\documentclass[%
superscriptaddress,
preprint,
nofootinbib,
 amsmath,amssymb,
 aps,
 longbibliography,
pra,
floatfix,
]{revtex4-1}

\usepackage[]{graphicx}% Include figure files
\usepackage{dcolumn}% Align table columns on decimal point
\usepackage{bm}% bold math

\usepackage{hyperref}% add hypertext capabilities
\hypersetup{
  final=true,
  plainpages=false,
  pdfstartview=FitV,
  pdftoolbar=true,
  pdfmenubar=true,
  bookmarksopen=true,
  bookmarksnumbered=true,
  breaklinks=true,
  linktocpage,
  colorlinks=true,
  linkcolor=black,
  urlcolor=blue,
  citecolor=black,
  anchorcolor=black
}

\usepackage{mathtools}

\DeclarePairedDelimiter\bra{\langle}{\rvert}
\DeclarePairedDelimiter\ket{\lvert}{\rangle}

\newenvironment{spmatrix}[1]
{\def\mysubscript{#1}\mathop\bgroup\begin{pmatrix}}
	{\end{pmatrix}\egroup_{\textstyle\mathstrut\mysubscript}}

\renewcommand\vec[1]{\mathbf{#1}}

\begin{document}

\title{Electrically driven electron spin resonance mediated by spin-valley-orbit coupling in a silicon quantum dot\footnote{Published on npj Quantum Information \textbf{4},6 (2018),  doi:\href{https://dx.doi.org/10.1038/s41534-018-0059-1}{10.1038/s41534-018-0059-1}. License \href{http://creativecommons.org/licenses/by/4.0/}{CC BY 4.0}}}
\author{Andrea Corna}
\affiliation{CEA, INAC-PHELIQS, 17 Rue des Martyrs, F-38000 Grenoble, France}
\affiliation{University Grenoble Alpes, Grenoble, France}
\author{L\'eo Bourdet}
\affiliation{CEA, INAC-MEM, 17 Rue des Martyrs, F-38000 Grenoble, France}
\affiliation{University Grenoble Alpes, Grenoble, France}
\author{Romain Maurand}
\affiliation{CEA, INAC-PHELIQS, 17 Rue des Martyrs, F-38000 Grenoble, France}
\affiliation{University Grenoble Alpes, Grenoble, France}
\author{Alessandro Crippa}
\affiliation{CEA, INAC-PHELIQS, 17 Rue des Martyrs, F-38000 Grenoble, France}
\affiliation{University Grenoble Alpes, Grenoble, France}
\author{Dharmraj Kotekar-Patil}
\affiliation{CEA, INAC-PHELIQS, 17 Rue des Martyrs, F-38000 Grenoble, France}
\affiliation{University Grenoble Alpes, Grenoble, France}
\author{Heorhii Bohuslavskyi}
\affiliation{CEA, INAC-PHELIQS, 17 Rue des Martyrs, F-38000 Grenoble, France}
\affiliation{CEA, LETI-MINATEC, 17 Rue des Martyrs, F-38000 Grenoble, France}
\affiliation{University Grenoble Alpes, Grenoble, France}
\author{Romain Lavi\'eville}
\affiliation{CEA, LETI-MINATEC, 17 Rue des Martyrs, F-38000 Grenoble, France}
\affiliation{University Grenoble Alpes, Grenoble, France}
\author{Louis Hutin}
\affiliation{CEA, LETI-MINATEC, 17 Rue des Martyrs, F-38000 Grenoble, France}
\affiliation{University Grenoble Alpes, Grenoble, France}
\author{Sylvain Barraud}
\affiliation{CEA, LETI-MINATEC, 17 Rue des Martyrs, F-38000 Grenoble, France}
\affiliation{University Grenoble Alpes, Grenoble, France}
\author{Xavier Jehl}
\affiliation{CEA, INAC-PHELIQS, 17 Rue des Martyrs, F-38000 Grenoble, France}
\affiliation{University Grenoble Alpes, Grenoble, France}
\author{Maud Vinet}
\affiliation{CEA, LETI-MINATEC, 17 Rue des Martyrs, F-38000 Grenoble, France}
\affiliation{University Grenoble Alpes, Grenoble, France}
\author{Silvano De Franceschi}
\affiliation{CEA, INAC-PHELIQS, 17 Rue des Martyrs, F-38000 Grenoble, France}
\affiliation{University Grenoble Alpes, Grenoble, France}
\author{Yann-Michel Niquet}
\affiliation{CEA, INAC-MEM, 17 Rue des Martyrs, F-38000 Grenoble, France}
\affiliation{University Grenoble Alpes, Grenoble, France}
\author{Marc Sanquer}
\affiliation{CEA, INAC-PHELIQS, 17 Rue des Martyrs, F-38000 Grenoble, France}
\affiliation{University Grenoble Alpes, Grenoble, France}

\makeatletter
\hypersetup{%
   pdfauthor={A. Corna et al.},
   pdftitle={\@title}
}

\makeatother

\maketitle

\textbf{The ability to manipulate electron spins with voltage-dependent electric f{}ields is key to the operation of quantum spintronics devices such as spin-based semiconductor qubits. A natural approach to electrical spin control exploits the spin-orbit coupling (SOC) inherently present in all materials. So far, this approach could not be applied to electrons in silicon, due to their extremely weak SOC. Here we report an experimental realization of electrically driven electron-spin resonance in a silicon-on-insulator (SOI) nanowire quantum dot device. The underlying driving mechanism results from an interplay between SOC and the multi-valley structure of the silicon conduction band, which is enhanced in the investigated nanowire geometry. We present a simple model capturing the essential physics and use tight-binding simulations for a more quantitative analysis. We discuss the relevance of our f{}indings to the development of compact and scalable electron-spin qubits in silicon.}

\section{Introduction}

Silicon is a strategic semiconductor for quantum spintronics, combining long spin coherence and mature technology \cite{Zwanenburg13}. Research on silicon-based spin qubits has seen a tremendous progress over the past f{}ive years. In particular, very long coherence times have been achieved with the introduction of devices based on the nuclear-spin-free $^{28}$Si isotope, enabling the suppression of hyperf{}ine coupling, the main source of spin decoherence \cite{Tyryshkin12}. Single qubits with f{}idelities exceeding 99\% as well as a f{}irst demonstration of a two-qubit gate have been reported \cite{Veldhorst14,Laucht15,Veldhorst15b}. 

f{}inding a viable pathway towards large-scale integration is the next step. To this aim, access to electric-f{}ield-mediated spin control would facilitate device scalability, circumventing the need for more demanding control schemes based on magnetic-f{}ield-driven spin resonance. Electric-f{}ield control requires a mechanism coupling spin and motional degrees of freedom. This so-called spin-orbit coupling (SOC) is generally present in atoms and solids -- due to a relativistic effect, electrons moving in an electric-f{}ield gradient experience in their reference frame an effective magnetic f{}ield. In the case of electrons in silicon, however, SOC is intrinsically very weak. 

Possible approaches to circumvent this limitation have so far relied either on the introduction of micromagnets, generating local magnetic f{}ield gradients and hence an artif{}icial SOC \cite{Pioro-Ladriere08,Kawakami14,Rancic16,Takeda16}, or on the use of hole spins \cite{Maurand16}, for which SOC is strong. In both cases, relatively fast coherent spin rotations could be achieved through resonant radio-frequency (RF) modulation of a control gate voltage. While the actual scalability of these two solutions remains to be investigated, other valuable opportunities may emerge from the rich physics of electrons in silicon nanostructures \cite{Nestoklon06,Zwanenburg13,Hao14,Schoenfield17,Jock17,Veldhorst15,Ruskov2017}.

Silicon is indeed an indirect band-gap semiconductor with six degenerate conduction-band valleys. This degeneracy is lifted in quantum dots (QD) where quantum conf{}inement leaves only two low-lying valleys that can be coupled by potential steps at Si/SiO$_2$ interfaces. The resulting valley eigenstates, which we label $v_1$ and $v_2$, are separated by an energy splitting $\Delta$ ranging from a few tens of $\mu$eV to a few meV \cite{Sham79,Saraiva09,Friesen10,Culcer10}. $\Delta$ depends on the conf{}inement potential, and can hence be tuned by externally applied electric f{}ields \cite{Goswami07,Yang13,Laucht15}. Even if weak \cite{Rashba08}, SOC can couple valley and spin degrees of freedom when, following the application of a magnetic f{}ield, $E_Z \sim \Delta$, where $E_Z$ is the Zeeman energy splitting. It has been shown that this operating regime can result in enhanced spin relaxation \cite{Yang13,Veldhorst14,Huang2014}.

Here we demonstrate that it can be exploited to perform electric-dipole spin resonance (EDSR) \cite{Golovach06,Nowack07,Nadj-Perge10,Pribiag13}. This functionality is enabled by the use of a quantum dot with a low-symmetry conf{}inement potential. We discuss the implications of these results for the development of silicon spin qubits. 

\section{Results}

\begin{figure}
\includegraphics[width=.85\textwidth]{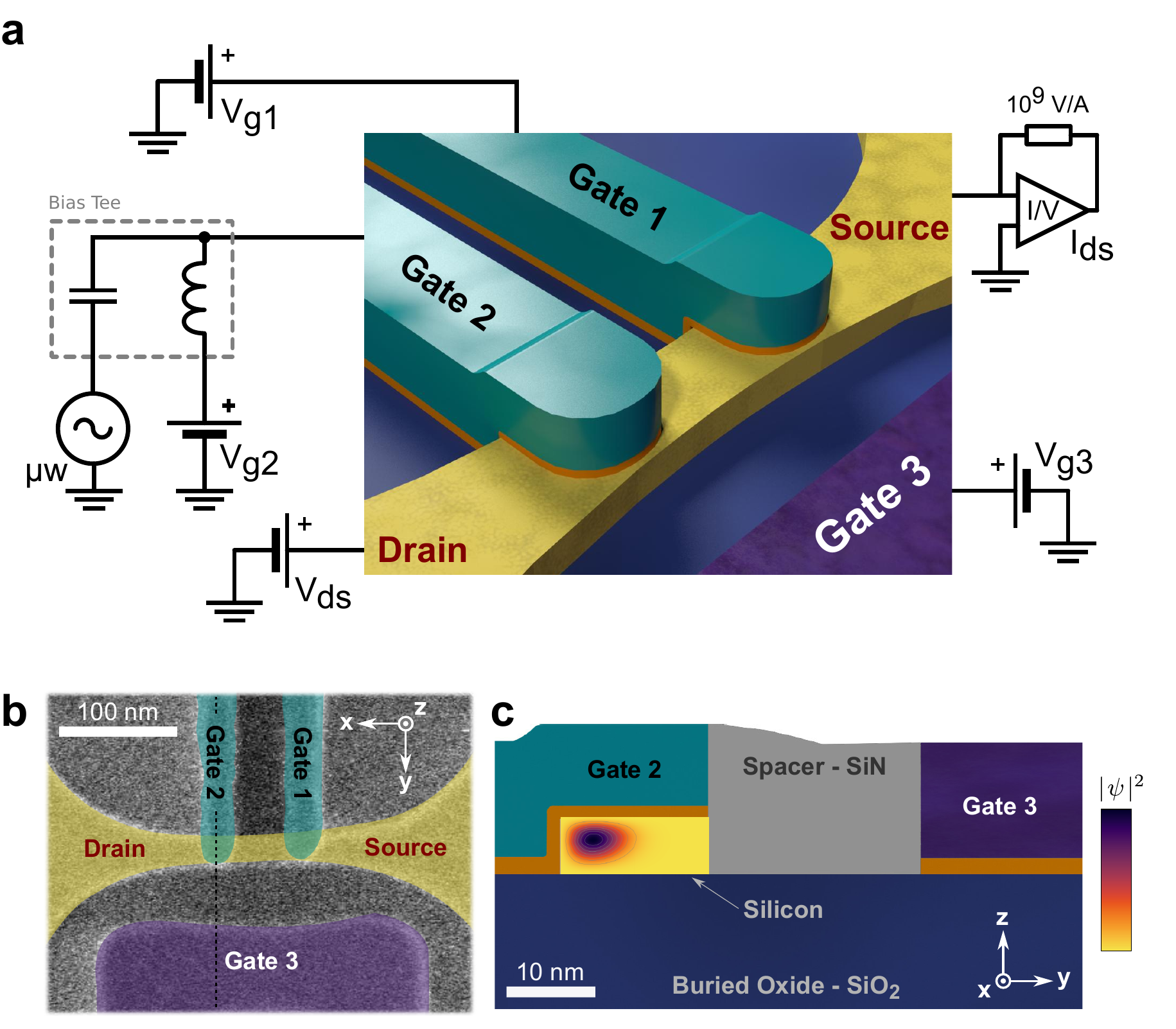}
\caption{\textbf{Device schematics.} \textbf{a}, Sketch of the sample and measurement setup. The silicon nanowire with the source/drain extensions is colored in yellow, the two top gates in green, and the side gate in violet. The gate oxides are colored in orange and the buried oxide (BOX) of the SOI in blue. \textbf{b}, Colorized device top view obtained by scanning electron microscopy before the deposition of the spacers of a device similar to the one used in the experiment. False-colors as in \textbf{a}. \textbf{c}, Sketch of the transverse cross-section of the device along the dashed line in \textbf{b}, with the spacer in gray. The cross-section of the silicon nanowire (yellow background) shows a color map of the square of the tight-binding wave function of the lowest conduction band state.}
\label{fig1}
\end{figure}

The experiment is carried out on a silicon nanowire device fabricated on a 300 mm diameter silicon-on-insulator wafer using an industrial-scale fabrication line \cite{Maurand16}. The device, shown in the schematic of f{}ig.~\ref{fig1}a and in the scanning electron micrograph of f{}ig.~\ref{fig1}b, consists of an undoped, $30$\,nm-wide and $12$\,nm-thick silicon channel oriented along $[110]$, with $n$-doped contacts. Two $35$\,nm-wide top-gates (gate\,1 and gate\,2), spaced by $30$\,nm, partially cover the channel. An additional gate (gate\,3) is located on the opposite side at a distance of $50$\,nm from the nanowire. Electron transport measurements were performed in a dilution refrigerator with a base temperature $T=15$\,mK. At this temperature, two QDs in series, labeled as QD1 and QD2, can be def{}ined by the accumulation voltages $V_{g1}$ and $V_{g2}$ applied to gate\,1 and gate\,2, respectively. The two QDs are conf{}ined against the nanowire edge covered by the gates, forming so-called ``corner'' dots \cite{Voisin14,Gonzalez15}, as conf{}irmed by tight-binding simulations of the lowest energy states, whose wave-functions are shown in f{}ig.~\ref{fig1}c. We tune the electron f{}illing of QD1 and QD2 down to relatively small occupation numbers $n_1$ and $n_2$, respectively ($n_1,n_2<10$, as inferred from the threshold voltage at room temperature and the charging energy \cite{Maurand16}). The side gate is set to a negative $V_{g3}=-0.28$\,V in order to further push the QD wave-functions against the opposite nanowire edges. 

In the limit of vanishing inter-dot coupling and odd occupation numbers, both QD1 and QD2 have a spin-1/2 ground state. At f{}inite magnetic f{}ield, $\vec{B}$, the respective spin degeneracies are lifted by the Zeeman energy $E_Z=g\mu_{\rm B}B$, where $\mu_{\rm B}$ is the Bohr magneton and $g$ is the Land\'e $g$-factor, which is close to the bare electron value ($g\simeq 2$) for electrons in silicon \cite{Feher59b}. In essence, our experiment consists in measuring electron transport through the double dot while driving EDSR in QD2. The polarized spin in QD1 acts as an effective ``spin f{}ilter'' regulating the current flow as a function of the spin admixture induced by EDSR in QD2. This Pauli blockade regime can be achieved only when the double dot is biased in a charge/spin conf{}iguration where inter-dot tunneling is forbidden by spin conservation \cite{Hanson07}. The simplest case involves the inter-dot charge transition $(n_1=1,n_2=1) \to (n_1=0,n_2=2)$, where one electron tunnels from QD1 into QD2. The two electrons may indeed form singlet (S) or triplet (T) states. While the singlet $S(1,1)$ and triplet $T(1,1)$ states are only weakly split by exchange interations and magnetic f{}ield and may both be loaded, the triplet $T(0,2)$ states remain typically out of reach because they must involve some orbital excitation of QD2. The system may hence be trapped for long times in the $T(1,1)$ states since tunneling from $T(1,1)$ to the $S(0,2)$ ground-state is forbidden by Pauli exclusion principle \cite{Hanson07}. This scenario can be generalized to the $(n_1,n_2) \to (n_1-1, n_2+1)$ transitions where $n_1$ and $n_2$ are odd integers. The current is strongly suppressed unless EDSR mixes $T(1,1)$ and $S(1,1)$ by rotating the spin in QD2.  

\begin{figure}
\includegraphics[width=.85\textwidth]{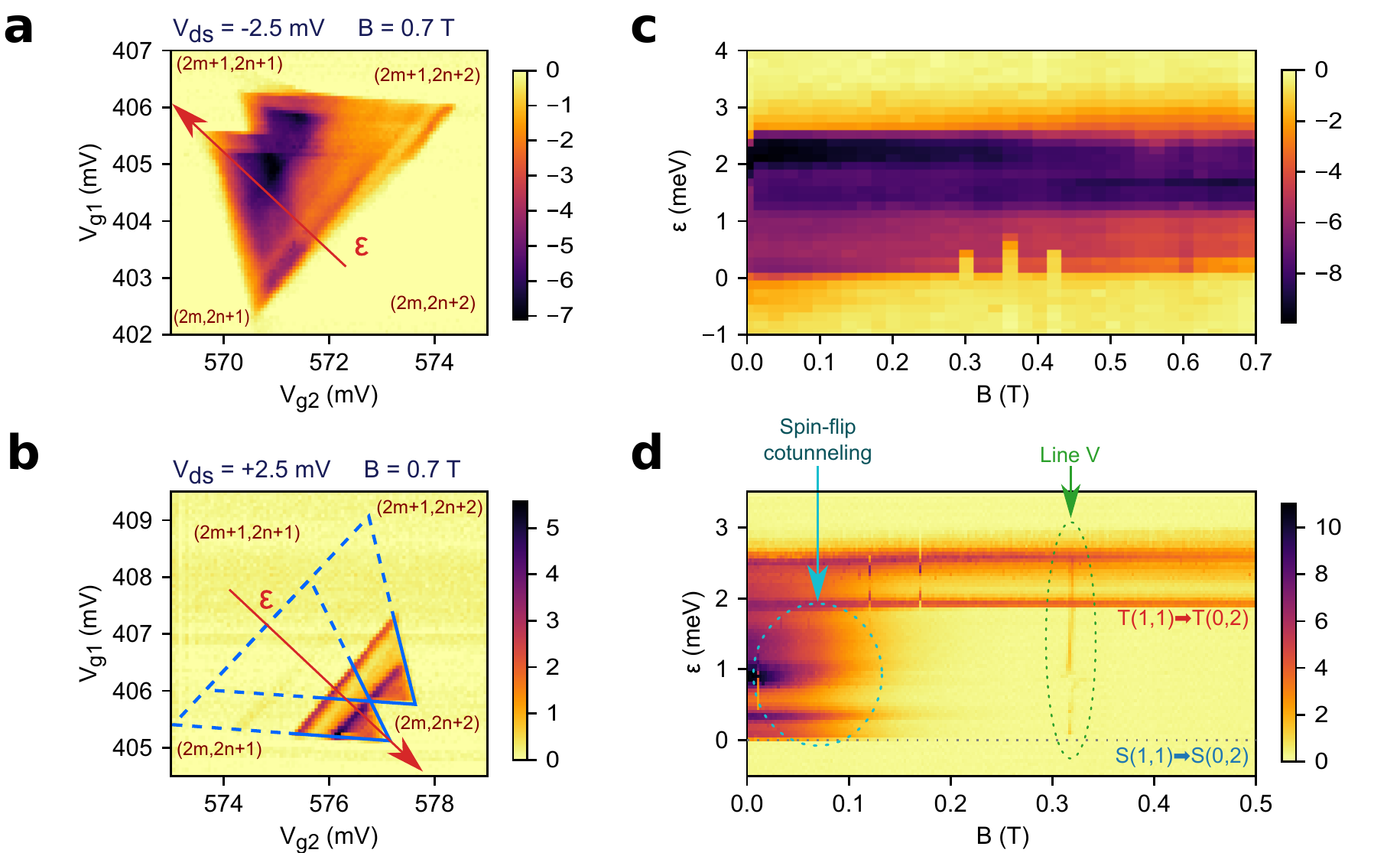}
\caption{\textbf{Pauli Spin Blockade.} \textbf{a}, Map of the current $I_{ds}$ as a function of $V_{g1}$ and $V_{g2}$ at f{}inite $V_{ds}=-2.5$ mV around a pair of triangles showing current rectif{}ication. The magnetic f{}ield $B=0.7$ T is parallel to $\vec{x}-\vec{y}$ (see axes on f{}ig.~\ref{fig1}) The number of electrons in each dot is given between parentheses, with $n$, $m$ integers. \textbf{b}, Same as \textbf{a} at opposite $V_{ds}=+2.5$ mV. \textbf{c}, Current as a function of detuning and magnetic f{}ield. The detuning axis is highlighted by a red arrow in \textbf{a}. \textbf{d}, Same as \textbf{c} at opposite $V_{ds}=+2.5$ mV with detuning axis def{}ined in \textbf{b}.}
\label{fig2}
\end{figure}

Because the opposite $(0,2) \to (1,1)$ transition (or, more generically, $(n_1-1,n_2+1) \to (n_1, n_2)$) is never blocked (there is always a (1,1) spin singlet to tunnel to), the Pauli blockade regime can be revealed by source-drain current rectif{}ication \cite{Ono02}. f{}ig.~\ref{fig2} presents measurements of the source-drain current, $I_{ds}$, as a function of $(V_{g1},V_{g2})$ in a charge conf{}iguration exhibiting Pauli rectif{}ication. f{}ig.~\ref{fig2}a corresponds to a source-drain bias voltage $V_{ds}=-2.5$\,mV and a magnetic f{}ield $B=0.7$\,T. Current flows within characteristic triangular regions \cite{Hanson07} where the electrochemical potential of dot 1, $\mu_1(n_1,n_2)$, is lower than the electrochemical potential of dot 2, $\mu_2(n_1-1,n_2+1)$. The energy detuning $\varepsilon$ between the two electrochemical potentials increases when moving along the red arrow. Current contains contributions from both elastic (i.e. resonant) and inelastic inter-dot tunneling. f{}ig.~\ref{fig2}b shows that reversing the bias voltage (i.e.  $V_{ds}=2.5$\,mV) yields the desired Pauli rectif{}ication characterized by truncated current triangles (In Supplementary Note 1 we discuss the presence of a concomitant valley-blockade effect similar to the one shown by Hao \textit{et al.} \cite{Hao14}).

The extent of the spin-blockade region measured along the detuning axis corresponds to the energy splitting, $\Delta_{\rm ST}$, between singlet and triplet states in the $(n_1-1, n_2+1)$ charge conf{}iguration (which is equivalent to $(0,2)$), basically the singlet-triplet splitting in QD2 \footnote{in the case of non-degenerate triplet states at f{}inite $B$, $\Delta_{\rm ST}$ is the splitting with respect to the triplet state, $T_0$, with zero spin projection along $\vec{B}$}. We f{}ind $\Delta_{\rm ST}=1.9$\,meV. f{}igures~\ref{fig2}-c) and ~\ref{fig2}-d) show $I_{ds}$ as a function of $B$ and $\varepsilon$ for negative and positive $V_{ds}$, respectively. As expected \cite{Ono02}, in the non-spin blocked polarity (f{}ig.~\ref{fig2}c) $I_{ds}$ shows essentially no dependence on $B$. In the opposite polarity, spin blockade is lifted at low f{}ield ($B \lesssim 0.1$ T), due to spin-flip cotunneling \cite{Lai11,Yamahata12}, as well as at $B = 0.31$ T. This unexpected feature will be discussed later.

\begin{figure}
\includegraphics[width=.70\textwidth]{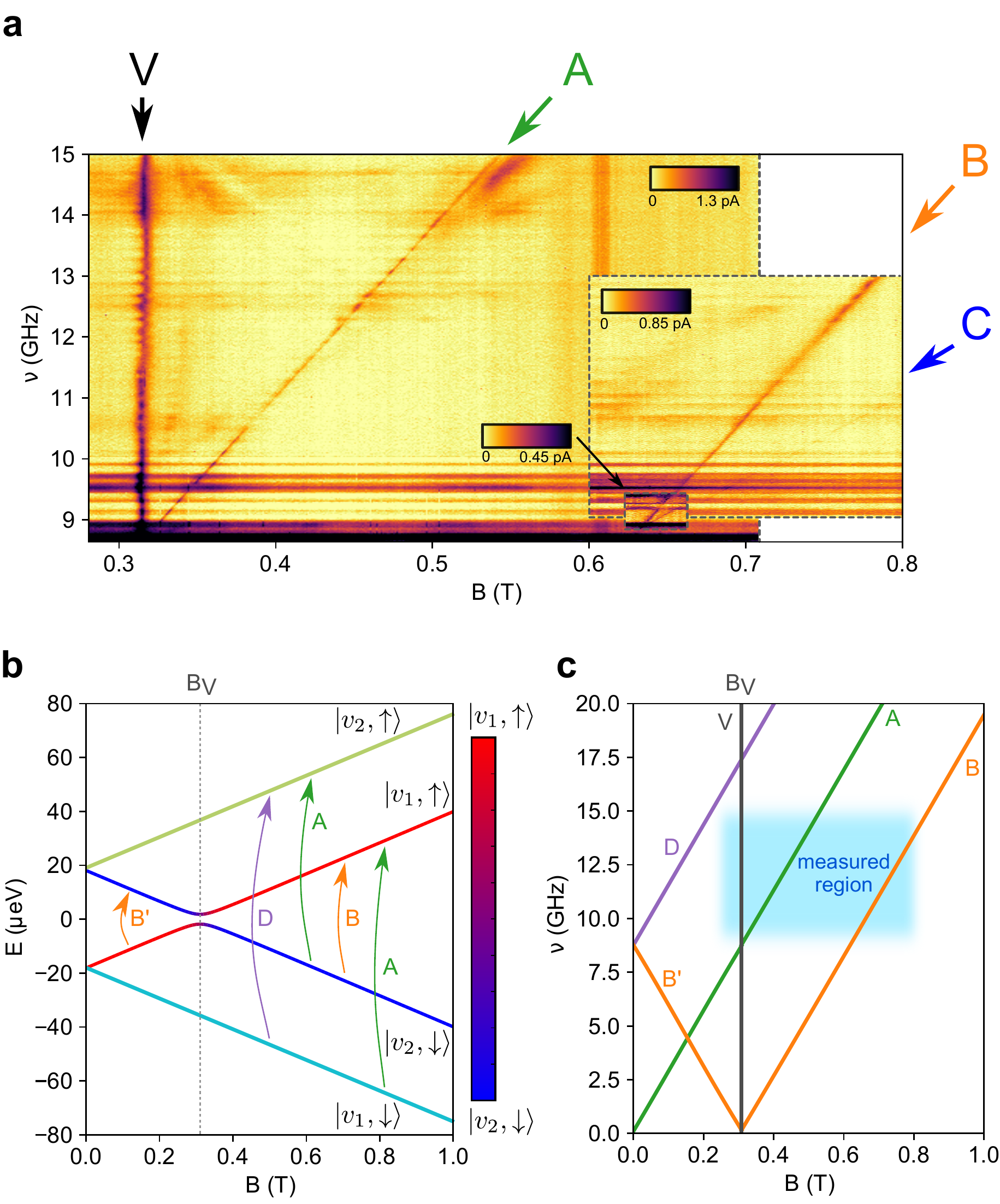}
\caption{\textbf{Spin Resonance.} \textbf{a}, Color plot of the measured drain current $I_{ds}$ as function of the magnetic f{}ield $B$ and microwave frequency $\nu$. The gates are biased in the spin-blockade regime. Three different measurements are gathered for clarity. EDSR transitions are revealed by oblique straight lines, labeled with letters A, B, and C. A vertical line (V) is also present. Line C is the weak feature that starts around $\approx 0.63$ T close to line B in the smallest inset. The faint vertical line at $\approx 0.61$T is an artifact. \textbf{b}, Diagram of the energy levels of a system with spin $S=1/2$ and two valleys ($v_1$, $v_2$) as function of magnetic f{}ield. At low magnetic f{}ield, the two lowest (highest) levels belongs to valley $v_1$ ($v_2$). Near the anti-crossing f{}ield $B_{\rm V}$, the states $\ket{v_1,\uparrow}$ and $\ket{v_2,\downarrow}$ hybridize due to SOC. The spin and valley composition of the hybridized states is quantif{}ied by the color scale on the right. The possible EDSR transitions are marked with green arrows (within $v_1$ or $v_2$) and orange/violet arrows (between $v_1$ and $v_2$). \textbf{c}, Sketch of the possible EDSR transitions in panel \textbf{b} as function of magnetic f{}ield and microwave frequency.}
\label{fig3}
\end{figure}

We now focus on the spin resonance experiment. To manipulate the spin electrically, we set $V_{g1}$ and $V_{g2}$ in the spin blocked region and apply a microwave excitation of frequency $\nu$ on gate 2. f{}ig.~\ref{fig3}a displays $I_{ds}$ as a function of $B$ and $\nu$ at constant power at the microwave source (The power at the sample depends on $\nu$ and is estimated to be $-53\,\text{dBm}\simeq0.7\,\text{mV}$ peak at $\nu\approx$ 9.6\,GHz). Several lines of increased current are visible in this plot, highlighting resonances along which Pauli spin blockade is lifted. They are labeled A, B, C and V. In the simplest case, spin resonance occurs when the microwave photon energy matches the Zeeman splitting between the two spin states of a doublet, i.e. when $h\nu=E_Z=g\mu_{\rm B}B$. We assign such a resonance to line A, because line A extrapolates to the origin ($B = 0$, $\nu = 0$). Its slope gives $g_{\rm A}=1.980 \pm 0.005$, which is compatible with the $g$-factor expected for electrons in silicon \cite{Feher59b}. Also line C extrapolates to the origin but with approximately half the slope (i.e. $g_{\rm C} = 0.96 \pm 0.01$). We attribute this line to a second-harmonic driving process \cite{Scarlino15}. 

We now focus on resonances B and V. The slope of line B is also compatible with the electron $g$-factor ($g_{\rm B}=2.00 \pm 0.01$). However, line B crosses zero-frequency at $B_{\rm V}=0.314 \pm 0.001$\,T, corresponding exactly to the magnetic f{}ield at which the non-dispersive resonance V appears. Consequently, line B can be assigned to transitions between spin states associated with two distinct orbitals. When these spin states cross at $B_{\rm V}$, Pauli spin blockade is lifted independently of the microwave excitation leading to the non-dispersive resonance V (see Supplementary Note 1 for details on the lifting of spin blockade at $B_{\rm V}$).

In order to understand the experimental EDSR spectrum of f{}ig.~\ref{fig3}a, we neglect in a f{}irst approximation the hybridization between the two QDs and consider only QD2 f{}illed with one electron. We have developed a model that accounts for the mixing between spin and valley states due to SOC. In our silicon nanowire geometry, the conf{}inement is strongest along the $z$ direction (normal to the SOI substrate), so that the low-energy levels belong to the $\Delta_{\pm z}$ valleys. Valley coupling at the Si/SiO$_2$ interface lifts the twofold valley degeneracy \cite{Sham79,Saraiva09,Friesen10,Culcer10,Zwanenburg13}, resulting in two spin-degenerate valley eigenstates $v_1$ and $v_2$ with energies $E_{1}$ and $E_{2}$, respectively, and a valley splitting $\Delta=E_{2}-E_{1}$. The expected energy diagram of the one-electron spin-valley states is plotted as a function of $B$ in f{}ig.~\ref{fig3}b (see Supplementary Note 2). The lowest spin-valley states can be identif{}ied as $\ket{v_1,\downarrow}$, $\ket{v_1,\uparrow}/\ket{v_2,\downarrow}$ anti-crossing at $B=B_{\rm V}$, and $\ket{v_2,\uparrow}$ (the spin being quantized along $\vec{B}$). From this energy diagram we assign the resonant transitions observed in f{}ig.~\ref{fig3}a as follows: line A corresponds to EDSR between states $\ket{v_1,\downarrow}$ and $\ket{v_1,\uparrow}$ (and between states $\ket{v_2,\downarrow}$ and $\ket{v_2,\uparrow}$); line B arises from EDSR between states $\ket{v_1,\uparrow}$ and $\ket{v_2,\downarrow}$ \cite{Scarlino17}; line V is associated with the anti-crossing between states $\ket{v_1,\uparrow}$ and $\ket{v_2,\downarrow}$ when $E_Z=\Delta$. We can thus measure $\Delta=g\mu_B B_V=36\,\mu$eV. Note that the experimental EDSR spectrum does not capture all possible transitions since some of them fall out of the scanned $(B,\nu)$ range. f{}ig.~\ref{fig3}c shows the expected EDSR spectrum starting from $\nu=0$ and $B=0$. The measured region is indicated in light blue (lower values of $\nu$ and $B$ could not be explored due to the onset of photon-assisted charge pumping and to the lifting of spin blockade, respectively).

The RF magnetic f{}ield associated with the microwave excitation on gate 2 is too weak to drive conventional ESR \cite{Golovach06}. Since a pure electric f{}ield cannot couple opposite spin states, SOC must be involved in the observed EDSR. The atomistic spin-orbit Hamiltonian primarily couples the different $p$ orbitals of silicon \cite{Chadi77}; the $\Delta_{\pm z}$ states are, however, linear combinations of $s$ and $p_z$ orbitals with little admixture of $p_x$ and $p_y$, which explains why the SOC matrix elements are weak in the conduction band of silicon. Yet the mixing between $\ket{v_1,\uparrow}$ and $\ket{v_2,\downarrow}$ by ``inter-valley'' SOC can be strongly enhanced when the splitting between these two states is small enough. We can capture the main physics and identify the relevant parameters using the simplest perturbation theory in the limit $B\ll B_{\rm V}$. The states $\ket{\Downarrow}\equiv\ket{v_1,\downarrow}$ and $\ket{\Uparrow}\equiv\ket{v_1,\uparrow}$ indeed read to f{}irst order in the spin-orbit Hamiltonian $H_{\rm SOC}$:
\begin{subequations}
\label{eqvpert}
\begin{align}
\ket{\Downarrow}&=\ket{v_1,\downarrow}-\frac{C_{v_1v_2}}{\Delta+g\mu_B B}\ket{v_2,\uparrow}+... \label{eqvperta} \\
\ket{\Uparrow}&=\ket{v_1,\uparrow}+\frac{C_{v_1v_2}^*}{\Delta-g\mu_B B}\ket{v_2,\downarrow}+... \,,
\end{align}
\end{subequations}
where: 
\begin{equation}
\label{eq:SOC}
C_{v_1v_2}=\bra{v_2,\uparrow}H_{\rm SOC}\ket{v_1,\downarrow}=-\bra{v_1,\uparrow}H_{\rm SOC}\ket{v_2,\downarrow}\,.
\end{equation}
Therefore, $\ket{v_1,\uparrow}$ admixes a signif{}icant fraction of $\ket{v_2,\downarrow}$ when the splitting $\Delta-g\mu_B B$ between these two states decreases. As $\ket{v_2,\downarrow}$ can be coupled to $\ket{v_1,\downarrow}$ by the RF electric f{}ield, this allows for Rabi oscillations between $\ket{\Uparrow}$ and $\ket{\Downarrow}$. Along line A, the Rabi frequency at resonance ($h\nu=g_A \mu_{\rm B}B$) reads:
\begin{equation}
hf=e\delta V_{g2}\left|\bra{\Uparrow}D\ket{\Downarrow}\right|\simeq 2eg\mu_B B\delta V_{g2}\frac{|D_{v_1v_2}||C_{v_1v_2}|}{\Delta^2}\,,
\label{eqrabipert}
\end{equation}
where $\delta V_{g2}$ is the amplitude of the microwave modulation on gate 2, $D(\textbf{r})=\partial{V_{t}(\textbf{r})}/\partial{V_{g2}}$ is the derivative of the total potential $V_{t}(\vec{r})$ in the device with respect to the gate potential $V_{g2}$, and:
\begin{equation}
D_{v_1v_2}=\bra{v_1,\uparrow}D\ket{v_2,\uparrow}=\bra{v_1,\downarrow}D\ket{v_2,\downarrow}
\end{equation}
is the matrix element of $D(\textbf{r})$ between valleys $v_1$ and $v_2$. The gate-induced electric f{}ield essentially drives motion in the $(yz)$ plane. $D_{v_1v_2}$ is small yet non negligible in SOI nanowire devices because the $v_1$ and $v_2$ wavefunctions show out of phase oscillations along $z$, and can hence be coupled by the vertical electric f{}ield. The f{}ield along $y$ does not result in a sizable $D_{v_1v_2}$ unless surface roughness disorder couples the motions along $z$ and in the $(xy)$ plane \cite{Gamble13,Boross16}. Although $C_{v_1v_2}$ is weak in silicon, SOC opens a path for an electrically driven spin resonance $\ket{\Downarrow}\to\ket{\Uparrow}$ through a virtual transition from $\ket{v_1,\downarrow}$ to $\ket{v_2,\downarrow}$, mediated by the microwave f{}ield, and then from $\ket{v_2,\downarrow}$ to $\ket{v_1,\uparrow}$, mediated by SOC. Note, however, that the above equations are only valid at small magnetic f{}ields where perturbation theory can be applied. A non-perturbative model valid at all f{}ields is introduced in the Supplementary Note 2. It explicitly accounts for the anti-crossing (and strong hybridization by SOC) of states $\ket{v_1,\uparrow}$ and $\ket{v_2,\downarrow}$ near $B=B_{\rm V}$ \cite{Yang13,Hao14,Huang2014}, but features the same matrix elements as above. This model shows that the Rabi frequency is maximal near $B=B_{\rm V}$, and conf{}irms that there is a concurrent spin resonance $\ket{v_1,\uparrow}\leftrightarrow\ket{v_2,\downarrow}$, shifted by the valley splitting $\Delta=36\,\mu$eV (lines B/B' on f{}ig.~\ref{fig3}c), as well as, in principle, a possible resonance $\ket{v_1,\downarrow}\leftrightarrow\ket{v_2,\uparrow}$ (line D).

\begin{figure}[ht]
	\includegraphics[width=.85\textwidth]{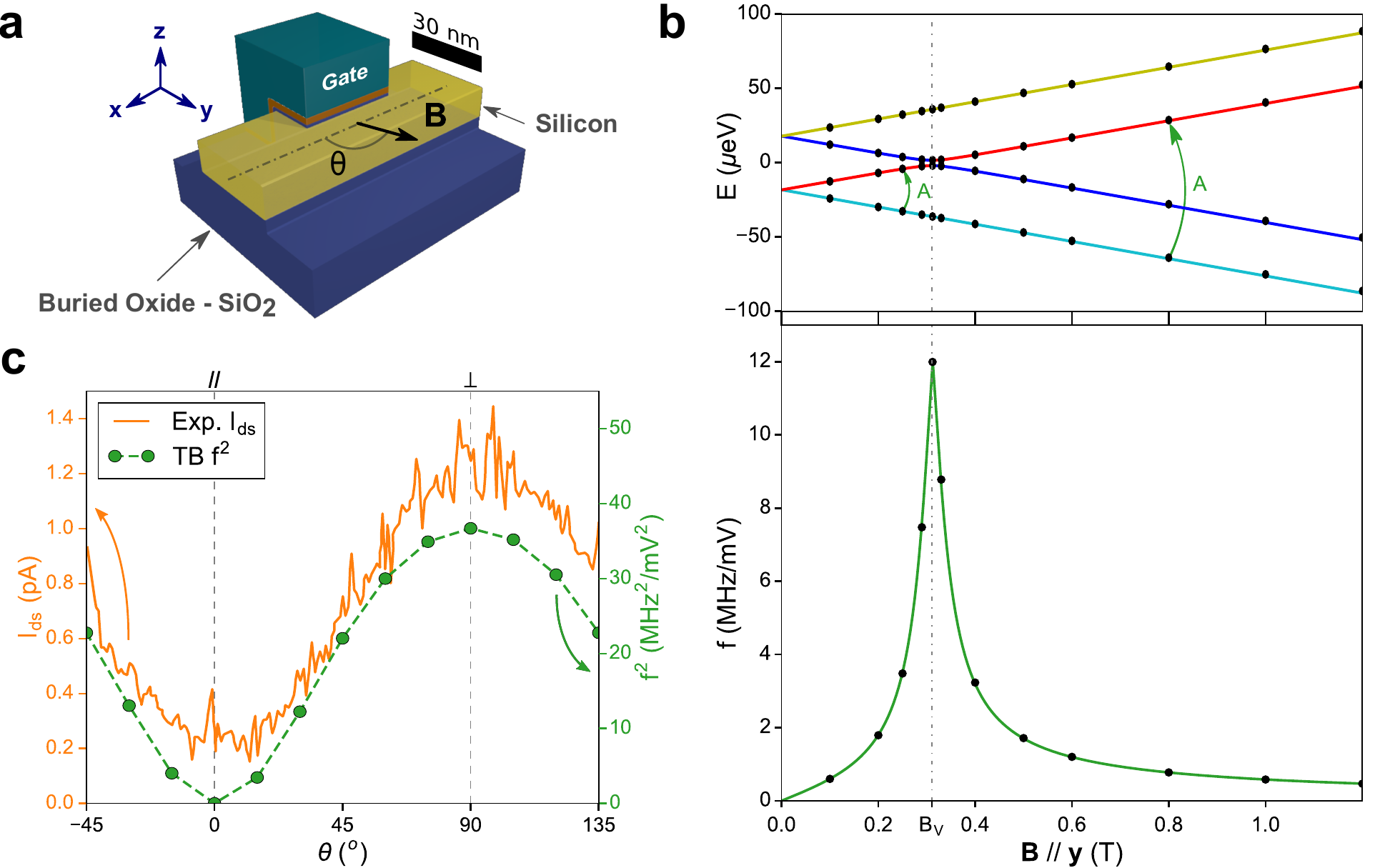}
	\caption{\textbf{TB calculations results.} \textbf{a}, Electrostatic model of the device. The silicon wire appears in yellow, SiO$_2$ in blue, HfO$_2$ in orange and the gate in green. We did not include the lateral gate in the simulations. \textbf{b}, TB energy levels as a function of the magnetic f{}ield $\vec{B}||\vec{y}$, and TB Rabi frequency along line A (identif{}ied by the green arrows) as a function of the magnetic f{}ield for a reference microwave amplitude $\delta V_{g2}=1$\,mV. The points are the TB data and the solid lines are the analytical model developed in Supplementary Note 3. The TB valley splitting energy $\Delta\sim36\,\mu$eV is in good agreement with the experimental value. The Rabi frequency shows a prominent peak near the anti-crossing f{}ield $B_{\rm V}$. \textbf{c}, Experimental measurement of EDSR current as a function of the angle $\theta$ between an in-plane magnetic f{}ield $\vec{B}\perp\vec{z}$ and the nanowire axis $\vec{x}$ ($\theta=0^\circ$), and TB Rabi frequency $f^2$ as a function of $\theta$.}
	\label{fig4}
\end{figure}

We have validated the above interpretation against $sp^3d^5s^*$ tight-binding (TB) calculations. TB is well suited to that purpose as it accounts for valley and spin-orbit coupling at the atomistic level. We consider a simplif{}ied single-gate device model capturing the essential geometry (f{}ig.~\ref{fig4}a). A realistic surface roughness disorder with rms amplitude $\Delta_{\rm SR}=0.4$ nm \cite{Bourdet16} is included in order to reduce the valley splitting down to the experimental value \cite{Culcer10}. A detailed description of the TB calculations is given in the Supplementary Note 3. The top panel of f{}ig.~\ref{fig4}b shows the $B$ dependence of the energy of the f{}irst four TB states $\ket{1}...\ket{4}$. An anti-crossing is visible between states $\ket{2}$ and $\ket{3}$ at $B_{\rm V}=0.3$\,T. We calculate $|C_{v_1v_2}|=1.8\,\mu$eV and $|D_{v_1v_2}|=70\,\mu$V/V. The $B$ dependence of the TB Rabi frequency $f$ on line A is shown in the bottom panel of f{}ig.~\ref{fig4}b. There is a prominent peak near $B=B_{\rm V}$ where $\ket{2}$ and $\ket{3}$ have a mixed $\ket{v_1,\uparrow}$/$\ket{v_2,\downarrow}$ character. The maximum Rabi frequency $f_{\rm max}\simeq e\delta V_{g2}|D_{v_1v_2}|/(\sqrt{2}h)$ is limited by $D_{v_1v_2}$ while the full width at half maximum of the peak, $\Delta B_{\rm FWHM}\simeq 12|C_{v_1v_2}|/(\sqrt{7}g\mu_B)=0.07$\,T is controlled by the SOC matrix element $C_{v_1v_2}$ (see Supplementary Note 3). The Rabi frequency remains however sizable over a few $\Delta B_{\rm FWHM}$. We point out that $C_{v_1v_2}$ and $D_{v_1v_2}$ may depend on the actual roughness at the Si/SiO$_2$ interface.

The calculated Rabi frequencies compare well against those reported for alternative silicon based systems. For example, the expected Rabi frequency is around 4.2 MHz for $B=0.35$ T, close to anticrossing f{}ield $B_{\rm V}$, and for a microwave excitation amplitude $\delta V_{g2}=0.7$\,mV, close to the experimental value (see f{}ig.~\ref{fig4}b). This frequency is comparable with those achieved with coplanar antennas \cite{Pla2012,Veldhorst14} and in some experiments with micromagnets \cite{Kawakami14,Takeda16}.

One of the most salient f{}ingerprint of the above EDSR mechanism is the dependence of $f$ on the magnetic f{}ield orientation. Indeed, it must be realized that $C_{v_1v_2}$ may vary with the orientation of the magnetic f{}ield (as the spin is quantized along $\vec{B}$ in Eq. (\ref{eq:SOC})). Actually, symmetry considerations supported by TB calculations show that $C_{v_1v_2}$ and hence $f$ are almost zero when $\vec{B}$ is aligned with the nanowire axis, due to the existence of a $(yz)$ mirror plane perpendicular to that axis (see Supplementary Note 4). As a simple hint of this result, we may consider a generic Rashba SOC Hamiltonian of the form $H_{\rm SOC}\propto(\vec{E}\times\vec{p})\cdot\boldsymbol{\sigma}$, where $\vec{E}$ is the electric f{}ield, $\vec{p}$ the momentum, and $\boldsymbol{\sigma}$ the Pauli matrices. Symmetric atoms on each side of the $(yz)$ plane contribute to $C_{v_1v_2}$ with opposite $E_x$ and $p_x$ components. Therefore, only the $\propto (E_yp_z-E_zp_y)\sigma_x$ component of $H_{\rm SOC}$ makes a non-zero contribution to $C_{v_1v_2}$, but does not couple opposite spins when $\vec{B}\parallel\vec{x}$. The current on line A is, to a f{}irst approximation, proportional to $f^2$ \cite{Koppens07,Schroer11}. The TB $f^2$ is plotted in f{}ig.~\ref{fig4}c as a function of the angle $\theta$ between an in-plane magnetic f{}ield $\vec{B}\perp\vec{z}$ and the nanowire axis $\vec{x}$. It shows the $\propto\sin^2\theta$ dependence expected from the above considerations. The experimental $I_{ds}$, also plotted in f{}ig.~\ref{fig4}c, shows the same behavior, supporting our interpretation. The fact that $I_{ds}$ remains f{}inite for $\vec{B}\parallel\vec{x}$ may be explained by the fact that the $(yz)$ symmetry plane is mildly broken by disorder and voltage biasing. 

In a recent work, Huang {\it et al.} \cite{Huang17} proposed a mechanism for EDSR based on electrically induced oscillations of an electron across an atomic step at a Si/SiO$_2$ or a Si/SiGe hetero-interface. The step enhances the SOC between the ground and the excited state of the same valley. The Rabi frequency is, however, limited by the maximal height of the step that the electron can overcome (typically 1 nm). The EDSR reported here has a different origin. It results from the f{}inite SOC and dipole matrix elements between the ground-states of valleys $v_1$ and $v_2$. These couplings are sizable only in a low-symmetry dot structure such as a corner QD \cite{Voisin14}, where there is only one mirror plane (see Supplementary Note 4). Indeed, we have verif{}ied that more symmetric device structures with at least two symmetry planes show a dramatic suppression of the SOC matrix element (since, as hinted above, each mirror rules out two out of three components in $H_{\rm SOC}$, leaving no possible coupling). This is the case of a typical nanowire f{}ield-effect transistor with the gate covering three sides of the nanowire channel, for which TB calculations give no EDSR since $C_{v_1v_2}=0$ whatever the spin or magnetic f{}ield orientation. Conversely, the design of QDs without any symmetry left should maximize the opportunities for EDSR.

\section{Discussion}

In conclusion, we have reported an experimental demonstration of electric-dipole, spin-valley resonance mediated by intrinsic SOC in a silicon electron double QD. Although SOC is weak in silicon, its effect can be enhanced in the corner QDs of an etched SOI device, owing to their reduced symmetry. SOC enables EDSR on the spin-split doublet of the f{}irst, lowest energy valley by mixing the up-spin state of that valley with the down-spin state of the second valley. The EDSR Rabi frequency is strongly enhanced near the corresponding anti-crossing, namely when the valley and Zeeman splittings are close enough. This enhancement comes with a price though, since we expect the spin relaxation time $T_1$ (and presumably also the spin coherence time $T_2$) to be simultaneously reduced \cite{Yang13,Huang2014}. Therefore, we anticipate that the eff{}iciency of the reported EDSR mechanism for spin qubit manipulation will be conditioned by the possibility to tune the valley splitting $\Delta$, in order to bring the qubit near the anti-crossing point for manipulation, then away from the anti-crossing point to mitigate decoherence. Given the strong dependence of the valley splitting on gate voltages in silicon-based devices \cite{Goswami07,Takashina2006}, this possibility appears within reach and will be addressed in future experiments.

\section*{Methods}
The silicon nanowire transistors are manufactured on a 300\,mm Silicon-On-Insulator (SOI) processing line \cite{Maurand16}. f{}irst, silicon nanowires are etched from a SOI wafer with a 12 nm\,thick undoped silicon layer and a 145\,nm thick buried oxide. The nanowire channels are oriented along the $[110]$ direction. The width $W$ of the nanowires, initially def{}ined by deep ultra-violet (DUV) lithography, is trimmed down to about 30\,nm by a well-controlled etching process. Two parallel top-gates, $\simeq 35$\,nm wide and spaced by $\simeq 30$\,nm are patterned with $e$-beam lithography in order to control the double quantum dot. An additional side gate is also placed parallel to the nanowire at a distance of 50\,nm in order to strengthen conf{}inement in the corner dots of the Si nanowire. The gate stack consists in a 2.5\,nm thick layer of SiO$_2$, a 1.9\,nm thick layer of HfO$_2$, a thin ($\simeq 5$\,nm) layer of TiN metal and a much thicker ($\simeq 50$\,nm) layer of polysilicon. Then, insulating SiN spacers are deposited all around the gates and are etched. Their width is deliberately large ($\simeq 25$\,nm) in order to cover completely the nanowire channel between the two gates and protect it from subsequent ion implantation. Arsenic and phosphorous are indeed implanted in order to achieve low resistance source/drain contacts. The wide spacers also limit dopant diffusion from the heavily implanted contact regions into the channel. The dopants are activated by spike annealing followed by silicidation. The devices are f{}inalized with a standard microelectronics back-end of line process.

The devices are f{}irst screened at room temperature. Those showing the best performances (symmetrical characteristics for both top gates with no gate leakage current, low subthreshold swing) are cleaved from the original 300-mm wafer in order to be mounted on a printed-circuit-board chip carrier with high-frequency lines. The sample is measured in a wet dilution fridge with a base temperature $T=15$\,mK. The magnetic f{}ield is applied by means of a 2D superconducting vector magnet in the $(xy)$ plane parallel to the SOI wafer.

All terminals are connected with bonding wires to DC lines; gate 2 is also connected with a bias-tee to a microwave line. The DC block is a low rise-time Tektronik PSPL5501A, while the RF f{}ilter is made with a 10\,k$\Omega$ SMD resistance mounted on the chip carrier plus a wire acting as inductor. The DC voltages are generated at room temperature by custom battery-powered opto-isolated voltage sources. The microwave signal is generated by a commercial analog microwave generator (Anritsu MG3693C). The RF line is equipped with a series of attenuators at room temperature and in the cryostat for signal thermalization ($1$\,K), with a total attenuation of $\simeq 38$\,dB at 10\,GHz. The current in the nanowire is measured by a custom transimpedance amplif{}ier with a gain of $10^9$\,V/A and then digitized by a commercial multimeter (Agilent 34410A).

The tight-binding calculations are performed with the $sp^3d^5s^*$ model of Ref. \onlinecite{Niquet09}. The potential in the device is calculated with a f{}inite-volume Poisson solver, then the eigenstates of the dot are computed with an iterative Jacobi-Davidson solver. The Rabi frequencies are obtained from Eq. (\ref{eqrabipert}).

\section*{Data availability}
The data that support the f{}indings of this study are available from the corresponding authors upon reasonable request.

\begin{acknowledgments}
This work was supported by the European Union's Horizon 2020 research and innovation program under grant agreement No 688539 MOSQUITO. Part of the calculations were run on the TGCC/Curie and CINECA/Marconi machines using allocations from GENCI and PRACE. We thank Cosimo Orban for the 3D rendering of the sample.
\end{acknowledgments}

\section*{Competing interests}
The authors declare no competing f{}inancial interests.

\section*{Author contributions}
H.B., R.L., L.H., S.B., and M.V. led device fabrication. A. Corna performed the experiment with help from R.M., A. Crippa and D.K.-P. under the supervision of X.J., S.D.F. and M.S. L. B. and Y.-M.N. did the modeling and simulations. M.V., X.J., S.D.F. and M.S led the all project. All authors co-wrote the manuscript.

\section*{Corresponding authors}
Correspondence and requests of material should be addressed to Marc Sanquer \\(\href{mailto:marc.sanquer@cea.fr}{marc.sanquer@cea.fr}), Silvano De Franceschi (\href{mailto:silvano.defranceschi@cea.fr}{silvano.defranceschi@cea.fr}) or Yann-Michel Niquet (\href{mailto:yann-michel.niquet@cea.fr}{yann-michel.niquet@cea.fr}).

%\section*{References}
%\bibliographystyle{naturemag}
%\bibliography{biblio_clean}

%%Supplementary
%%%%%%%%%% Merge with supplemental materials %%%%%%%%%%
\pagebreak
\widetext
\begin{center}
	\textbf{\large Supplementary notes for ``Electrically driven electron spin resonance mediated by spin-valley-orbit coupling in a silicon quantum dot''}
\end{center}
%%%%%%%%%% Merge with supplemental materials %%%%%%%%%%
%%%%%%%%%% Prefix a "S" to all equations, figures, tables and reset the counter %%%%%%%%%%
\setcounter{equation}{0}
\setcounter{figure}{0}
\setcounter{table}{0}
\setcounter{page}{1}
\makeatletter
\renewcommand{\theequation}{S\arabic{equation}}
\renewcommand{\thefigure}{S\arabic{figure}}
\renewcommand{\theHfigure}{S\arabic{figure}}
\renewcommand{\bibnumfmt}[1]{[S#1]}
\renewcommand{\citenumfont}[1]{S#1}
\renewcommand{\thepage}{S\arabic{page}}
%%%%%%%%%% Prefix a "S" to all equations, figures, tables and reset the counter %%%%%%%%%%

\section*{Supplementary note 1: Valleys and spin-valley blockade}
\label{par:spin-valley-block}

In the main text, we have discussed the nature of the A, B, C and V line in a one-particle picture. In this supplementary note, we introduce a two-particle picture for the blockade, which accounts for the valley degree of freedom and gives a better description of the V line.

Spin blockade can arise when the current flows through the sequence of charge configurations $(n_1,n_2)\equiv(2n+1, 2m+1)\to(2n, 2m+2)\to(2n, 2m+1)\to(2n+1, 2m+1)...$, where $n_1$ and $n_2$ are the number of electrons in dots 1 and 2.\cite{SOno02,SHanson07} Indeed, the $(2n+1, 2m+1)$ states can be mapped onto singlet $S(1, 1)$ and triplet $T(1, 1)$ states, while the $(2n, 2m+2)$ states can be mapped onto singlet $S(0, 2)$ and triplet $T(0, 2)$ states. While the $S(1, 1)$ and $T(1, 1)$ states are almost degenerate, the $S(0, 2)$ and $T(0, 2)$ states can be significantly split because the $T(0, 2)$ state must involve some orbital excitation. The $(2n+1)^{\rm th}$ electron may enter in dot 1 through any of the $(1, 1)$ configurations at high enough source-drain bias. Once in a $T(1, 1)$ state, the system may, however, get trapped for a long time if the $T(0, 2)$ states are still out of the bias window, because tunneling from $T(1, 1)$ to $S(0, 2)$ requires a spin flip. The current is hence suppressed. At reverse source-drain bias, the current flows through the sequence of charge configurations $(2n+1, 2m+1)\to(2n, 2m+1)\to(2n, 2m+2)\to(2n+1, 2m+1)...$, which can not be spin-blocked, giving rise to current rectification (see Fig. 2 of main text).

The observation of inter-valleys resonances suggests that $m$ is even (otherwise only transitions between $v_2$ states would be observed in dot 2). We assume from now on that $n$ is also even. As a matter of fact, the absence of {\it visible} bias triangles for lower gate voltages suggests that $m=n=0$, though we can not exclude the existence of extra triangles with currents below the detection limit.

\begin{figure}
	\centering
	\includegraphics[width=1.0\textwidth, angle=0]{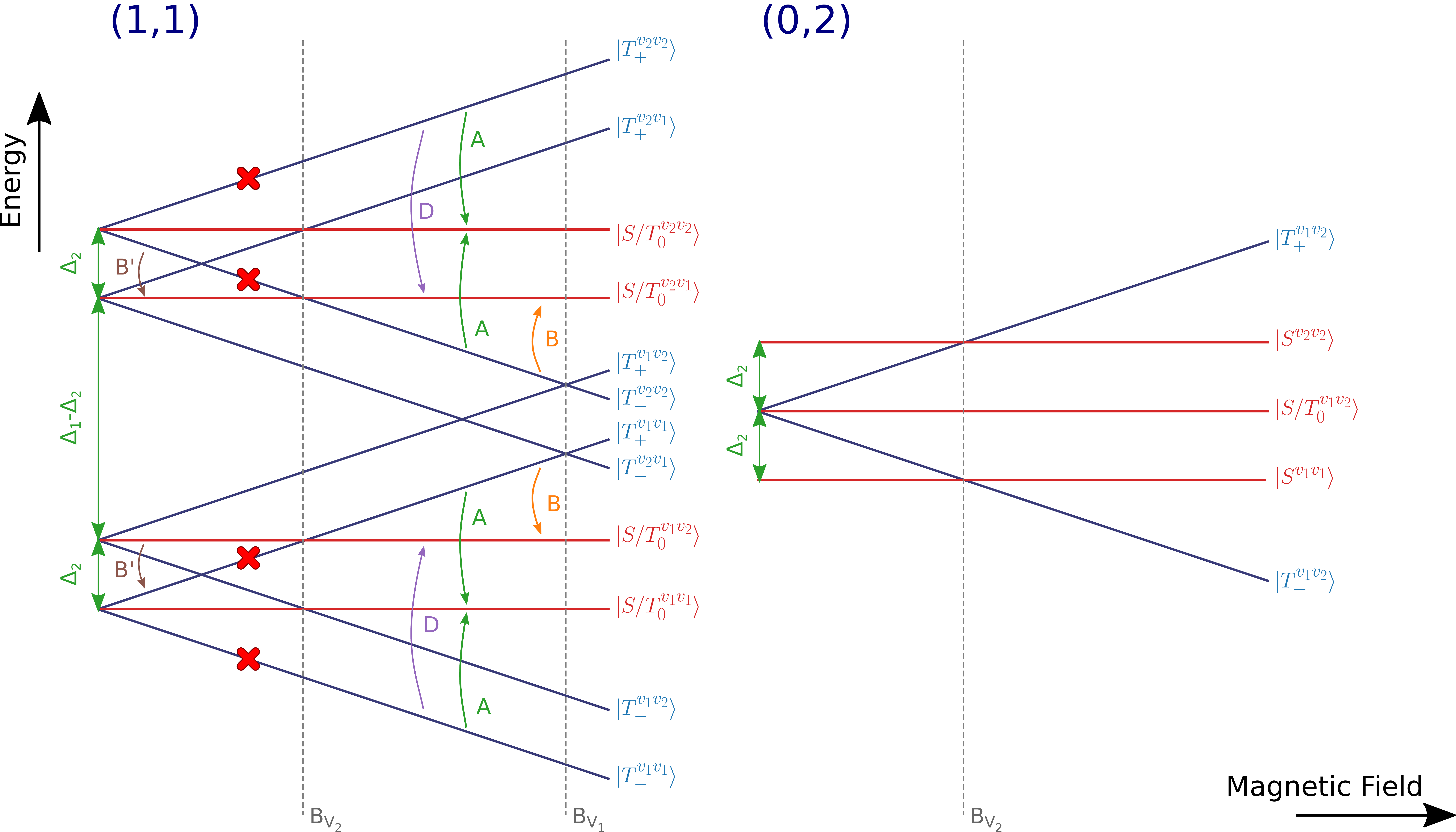}
	\caption{Scheme of the energy levels for a system with two dots, filled with two electrons in the $(1, 1)$ and $(0, 2)$ charge configurations, as a function of magnetic field. We assume different valley splittings in the two dots (respectively $\Delta_1>\Delta_2$). Triplets in the $(0, 2)$ configurations with two electrons in the same valley are not represented, since they are energetically far away. Spin-valley blocked states are marked with a red cross.}
	\label{fig:spin_valley_blockade}
\end{figure}

We discuss below the role of valley blockade in the present experiments. We assume the valley splitting is much larger in dot 1 ($\Delta_1$) than in dot 2 ($\Delta_2=36\,\mu$eV) due to disorder and bias conditions. The valley splitting in dot 1 is actually beyond the bandwidth of the EDSR setup. %, but transport measurements suggest that it could be as large as 350 $\mu$eV. 
This reflects the stochastic variations from one dot to an other, as confirmed by tight-binding simulations.

In the $(1, 1)$ charge configuration, the low-energy states can be characterized by their spin component [singlet (S) or triplet (T$_0$, T$_-$, T$_+$)] and by the valley occupied in each dot ($v_1$ or $v_2$). Sixteen states can be constructed in this way (see Fig. \ref{fig:spin_valley_blockade}). We neglect in a first approximation the small exchange splitting between singlet and triplet states with same valley indices. The magnetic field $\vec{B}=B\vec{b}$ splits the T$_-$ (total spin $\langle S_\vec{b}\rangle=-1$) and T$_+$ states ($\langle S_\vec{b}\rangle=+1$) from the S and $T_0$ states ($\langle S_\vec{b}\rangle=0$). The splitting between T$_+$ and T$_-$ is $E_z=2g\mu_BB$.

Similar states can be constructed in the $(0, 2)$ charge configuration. The $S^{v_1v_1}(0, 2)$ and $S^{v_1v_1}(1, 1)$ are detuned by the bias on gates 1 and 2. We focus on detunings smaller than the orbital singlet-triplet splitting $\Delta_{\rm ST}=1.9$ meV, so that neither the $v_1v_1$ nor the $v_2v_2$ triplets can be reached from the $(1, 1)$ states.

Given the small $\Delta_2=36\,\mu$eV extracted from spin resonance, we need, however, to reconsider the mechanisms for current rectification. Indeed, the system must be spin-and-valley blocked\cite{SHao14} since the detuning is typically much larger than $\Delta_2$ so that $T^{v_1v_2}(0, 2)$ states are accessible in the bias window. Assuming that both spin and valley are conserved during tunneling, the spin and valley blocked $(1, 1)$ states are actually $T_-^{v_1v_1}$, $T_+^{v_1v_1}$, $T_-^{v_2v_2}$ and $T_+^{v_2v_2}$. Although $T_0^{v_1v_1}$ and $T_0^{v_2v_2}$ are, in principle, also spin and valley blocked, they may be mixed with the nearly degenerate $S^{v_1v_1}$ and $S^{v_2v_2}$ states by, e.g., spin-orbit coupling (SOC) and nuclear spin disorder,\cite{SKoppens05,SNadj-Perge10b} and be therefore practically unblocked.

We can now refine the interpretation of the different lines. Each one corresponds to a different set of transitions between blocked and unblocked $(1, 1)$ states. Line A corresponds to transitions between $T_\pm^{v_1v_1}$ and $T_0^{v_1v_1}$/$S^{v_1v_1}$ states, and to transitions between $T_\pm^{v_2v_2}$ and $T_0^{v_2v_2}$/$S^{v_2v_2}$ states. Line B corresponds to transitions between $T_+^{v_1v_1}$ and $T_0^{v_1v_2}$/$S^{v_1v_2}$ states, and between $T_-^{v_2v_2}$ and $T_0^{v_2v_1}$/$S^{v_2v_1}$ states. The line D on Fig. 3 of the main text would correspond to transitions between $T_-^{v_1v_1}$ and $T_0^{v_1v_2}$/$S^{v_1v_2}$ states, and between $T_+^{v_2v_2}$ and $T_0^{v_2v_1}$/$S^{v_2v_1}$ states. These transitions give rise to the same spectrum as in the one-particle picture.  

Line V is independent on the microwave frequency and also appears when no microwaves are applied. At the magnetic field $B_{\rm V}\simeq\Delta_2/(g\mu_B)$, the states $T_+^{v_1v_1}$ and $T_0^{v_1v_2}/S^{v_1v_2}$, as well as the states $T_-^{v_2v_2}$ and $T_0^{v_2v_1}/S^{v_2v_1}$ are almost degenerate. The mixing of these near degenerate blocked and unblocked states by SOC lifts spin and valley blockade of the $T_+^{v_1v_1}$ and $T_-^{v_2v_2}$ states, giving rise to an excess of current at $B\simeq B_{\rm V}$ independent on the microwave excitation.

Note that we also expect a horizontal line H (independent on the magnetic field) corresponding to transitions between $T_\pm^{v_1v_1}$ and $T_\pm^{v_1v_2}$, and between $T_\pm^{v_2v_1}$ and $T_\pm^{v_2v_2}$, at frequency $\nu=\Delta/h=8.7$ GHz. Horizonal lines are indeed visible in the expected frequency range on Fig. 3a (main text), but they can not be unequivocally assigned due to the onset of photon-assisted charge pumping resulting in parasitic features.

\newpage
\section*{Supplementary note 2: Theory}
\label{sectionEDSR}

In this supplementary note, we propose a model for spin-orbit driven EDSR in silicon quantum dots accounting for the strong hybridization of the spin and valley states near the anti-crossing field $B=B_{\rm V}$.

We consider a silicon QD with strongest confinement along the $z$ direction so that the low-lying conduction band levels belong to the $\Delta_{\pm z}$ valleys. This dot is controlled by a gate with potential $V_g$. In the absence of valley and spin-orbit coupling, the ground-state is fourfold degenerate (twice for spins and twice for valleys). Valley coupling\cite{SSham79,SSaraiva09,SFriesen10,SCulcer10,SZwanenburg13} splits this fourfold degenerate level into two spin-degenerate states $\ket{v_1,\sigma}$ and $\ket{v_2,\sigma}$ with energies $E_1$ and $E_2$, separated by the valley splitting energy $\Delta=E_2-E_1$ ($\sigma=\,\uparrow,\downarrow$ is the spin index). In the simplest approximation, $\ket{v_1,\sigma}$ and $\ket{v_2,\sigma}$ are bonding and anti-bonding combinations of the $\Delta_{\pm z}$ states.

The remaining spin degeneracy can be lifted by a static magnetic field $\vec{B}$. The energy of state $\ket{v_n,\sigma}$ is then $E_{n,\sigma}=E_n\pm\frac{1}{2}g\mu_B B$ ($+$ for up states, $-$ for down states, the spin being quantized along $\vec{B}$). Here $g$ is the gyro-magnetic factor of the electrons, which is expected to be close to $g_0=2.0023$ in silicon.\cite{SFeher59b} We may neglect the effects of the magnetic field on the orbital motion of the electrons in a first approximation. The wave functions $\varphi_{n,\sigma}(\vec{r})=\langle\vec{r}|v_n,\sigma\rangle$ can then be chosen real. 

The gate potential is modulated by a RF signal with frequency $\nu$ and amplitude $\delta V_g$ in order to drive EDSR between states $\ket{v_1,\downarrow}$ and $\ket{v_1,\uparrow}$. At resonance $h\nu=g\mu_B B$, the Rabi frequency reads:
\begin{equation}
hf=e\delta V_g\left|\bra{v_1,\uparrow}D\ket{v_1,\downarrow}\right|\,,
\label{eqrabi}
\end{equation}
where $D(\vec{r})=\partial V_t(\vec{r})/\partial V_g$ is the derivative of the total potential $V_t(\vec{r})$ in the device with respect to $V_g$.\footnote{If the electrostatics of the device is linear, $D(\vec{r})=V_t^0(\vec{r})/V_g$, where $V_t^0(\vec{r})$ is the total potential in the empty device with the gate at potential $V_g$, and all other terminals grounded.} We discard the effects of the displacement currents (concomitant ESR), which are negligible.\cite{SGolovach06} In the absence of SOC, $hf$ is zero as an electric field can not couple opposite spins.

As discussed in the main text, spin-orbit couples the orbital and spin motions of the electron.\cite{SNestoklon06,SYang13,SHuang2014,SHao14,SVeldhorst15,SHuang17} ``Intra-valley'' SOC mixes spins within the $\Delta_{+z}$ or the $\Delta_{-z}$ valley, while ``inter-valley'' SOC mixes spins between the $\Delta_{+z}$ and $\Delta_{-z}$ valleys. Both intra-valley and inter-valley SOC may couple $\ket{v_1,\sigma}$ with the excited states of valley 1. Inter-valley SOC  may also couple $\ket{v_1,\sigma}$ with all $v_2$ states. Its effects can be strongly enhanced if the valley splitting $\Delta$ is small enough. Indeed, in the simplest non-degenerate perturbation theory, the states $\ket{\Downarrow}\equiv\ket{v_1,\downarrow}$ and $\ket{\Uparrow}\equiv\ket{v_1,\uparrow}$ read to first order in the spin-orbit Hamiltonian $H_{\rm SOC}$:
\begin{subequations}
	\label{Seqvpert}
	\begin{align}
	\ket{\Downarrow}&=\ket{v_1,\downarrow}-\frac{C_{v_1v_2}}{\Delta+g\mu_B B}\ket{v_2,\uparrow}-\frac{iR_{v_1v_2}}{\Delta}\ket{v_2,\downarrow} \label{Seqvperta} \\
	\ket{\Uparrow}&=\ket{v_1,\uparrow}+\frac{C_{v_1v_2}^*}{\Delta-g\mu_B B}\ket{v_2,\downarrow}+\frac{iR_{v_1v_2}}{\Delta}\ket{v_2,\uparrow}\,,
	\end{align}
\end{subequations}
with
\begin{subequations}
	\begin{align}
	C_{v_1v_2}&=\bra{v_2,\uparrow}H_{\rm SOC}\ket{v_1,\downarrow}=-\bra{v_1,\uparrow}H_{\rm SOC}\ket{v_2,\downarrow} \\
	R_{v_1v_2}&=-i\bra{v_2,\downarrow}H_{\rm SOC}\ket{v_1,\downarrow}=-i\bra{v_1,\uparrow}H_{\rm SOC}\ket{v_2,\uparrow}\,.
	\end{align}
\end{subequations}
The above equalities follow from time-reversal symmetry considerations for real wave functions. $C_{v_1v_2}$ is complex and $R_{v_1v_2}$ is real. We have neglected all mixing beyond the four $\ket{v_n,\sigma}$, because the higher-lying excited states usually lie $\gtrsim 1$ meV above $E_1$ and $E_2$ (as inferred from $\Delta_{\rm ST}$ on Fig. 2, main text). 

Inserting Eqs. (\ref{Seqvpert}) into Eq. (\ref{eqrabi}), then expanding in powers of $B$ yields to first order in $B$ and $H_{\rm SOC}$:
\begin{equation}
hf=e\delta V_g\left|\bra{\Uparrow}D\ket{\Downarrow}\right|=2eg\mu_B B\delta V_g\frac{|C_{v_1v_2}||D_{v_1v_2}|}{\Delta^2}\,.
\label{Seqrabipert}
\end{equation}
where
\begin{equation}
D_{v_1v_2}=\bra{v_1,\sigma}D\ket{v_2,\sigma}=\bra{v_2,\sigma}D\ket{v_1,\sigma}
\end{equation}
is the electric dipole matrix element between valleys $v_1$ and $v_2$. As expected, the Rabi frequency is proportional to $\delta V_g D_{v_1v_2}$, $C_{v_1v_2}$, and to $B$ (as the contributions from the $\propto C_{v_1v_2}$ terms in Eqs. (\ref{Seqvpert}) cancel out if time-reversal symmetry is not broken by the $\propto g\mu_B B$ terms of the denominators). It is also inversely proportional to $\Delta^2$ ; namely the smaller the valley splitting, the faster the rotation of the spin. $R_{v_1v_2}$ does not contribute to lowest order because it couples states with the same spin. 

$C_{v_1v_2}$ and $D_{v_1v_2}$ are known to be small in the conduction band of silicon.\cite{SNestoklon06,SYang13,SHao14,SVeldhorst15,SHuang17} Actually, $D_{v_1v_2}$ is zero in any approximation that completely decouples the $\Delta_{\pm z}$ valleys (such as the simplest effective mass approximation). It is, however, finite in tight-binding\cite{SDiCarlo03,SDelerue05} or advanced $\vec{k}\cdot\vec{p}$ models\cite{SSaraiva09,SCulcer10} for the conduction band of silicon. According to Eq. (\ref{Seqrabipert}), the Rabi frequency can be significant if $E_{1,\uparrow}$ is close enough to $E_{2,\downarrow}$ to enhance spin-valley mixing by $H_{\rm SOC}$. This happens when $\Delta$ is small and/or when $g\mu_B B\approx\Delta$ (see later discussion). The main path for EDSR is then the virtual transition from $\ket{v_1,\downarrow}$ to $\ket{v_2,\uparrow}$ (mediated by $H_{\rm SOC}$), then from $\ket{v_2,\uparrow}$ to $\ket{v_1,\uparrow}$ (mediated by the RF field). 

The above equations are valid only for very small magnetic fields $B$, as non-degenerate perturbation theory breaks down near the anti-crossing between $E_{1,\uparrow}$ and $E_{2,\downarrow}$ when $g\mu_B B\approx\Delta$ (see Fig. 3b of the main text). We may deal with this anti-crossing using degenerate perturbation theory in the $\{\ket{v_1,\uparrow},\ket{v_2,\downarrow}\}$ subspace, while still using Eq. (\ref{Seqvperta}) for state $\ket{\Downarrow}$. However, such a strategy would spoil the cancellations between $\ket{\Downarrow}$ and $\ket{\Uparrow}$ needed to achieve the proper behavior $hf\to 0$ when $B\to 0$. We must, therefore, treat SOC in the full $\{\ket{v_1,\downarrow},\ket{v_1,\uparrow},\ket{v_2,\downarrow},\ket{v_2,\uparrow}\}$ subspace.\cite{SYang13,SHuang2014}

The total Hamiltonian then reads:
\begin{equation}
H=\begin{spmatrix}{}
E_1-\frac{1}{2}g\mu_B B & 0 & -iR_{v_1v_2} & C_{v_1v_2}^* \\
0 & E_1+\frac{1}{2}g\mu_B B & -C_{v_1v_2} & iR_{v_1v_2} \\
iR_{v_1v_2} & -C_{v_1v_2}^* & E_2-\frac{1}{2}g\mu_B B & 0 \\
C_{v_1v_2} & -iR_{v_1v_2} & 0 & E_2+\frac{1}{2}g\mu_B B \\
\end{spmatrix}\,.
\end{equation}
As discussed before, $R_{v_1v_2}$ is not expected to make significant contributions to the EDSR as it mixes states with the same spin. We may therefore set $R_{v_1v_2}=0$ for practical purposes; $H$ then splits into two $2\times2$ blocks in the $\{\ket{v_1,\uparrow},\ket{v_2,\downarrow}\}$ and $\{\ket{v_1,\downarrow},\ket{v_2,\uparrow}\}$ subspaces. The diagonalization of the $\{\ket{v_1,\downarrow},\ket{v_2,\uparrow}\}$ block yields energies:
\begin{equation}
E_\pm=\frac{1}{2}(E_1+E_2)\pm\frac{1}{2}\sqrt{(\Delta+g\mu_B B)^2+4|C_{v_1v_2}|^2}\,
\end{equation}
and eigenstates:
\begin{subequations}
	\begin{align}
	\ket{\psi_+}&=\alpha\ket{v_1,\downarrow}+\beta\ket{v_2,\uparrow} \\
	\ket{\psi_-}&=\beta\ket{v_1,\downarrow}-\alpha^*\ket{v_2,\uparrow}
	\end{align}
\end{subequations}
with:
\begin{subequations}
	\begin{align}
	\alpha&=\frac{-2C_{v_1v_2}}{\left(4|C_{v_1v_2}|^2+W^2\right)^{1/2}} \\
	\beta&=\frac{W}{\left(4|C_{v_1v_2}|^2+W^2\right)^{1/2}}
	\end{align}
\end{subequations}
and:
\begin{equation}
W=\Delta+g\mu_B B+\sqrt{(\Delta+g\mu_B B)^2+4|C_{v_1v_2}|^2}\,.
\end{equation}
Likewise, the diagonalization of the $\{\ket{v_1,\uparrow},\ket{v_2,\downarrow}\}$ block yields energies:
\begin{equation}
E_\pm^\prime=\frac{1}{2}(E_1+E_2)\pm\frac{1}{2}\sqrt{(\Delta-g\mu_B B)^2+4|C_{v_1v_2}|^2}\,
\end{equation}
and eigenstates:
\begin{subequations}
	\begin{align}
	\ket{\psi_+^\prime}&=\alpha^\prime\ket{v_1,\uparrow}+\beta^\prime\ket{v_2,\downarrow} \\
	\ket{\psi_-^\prime}&=\beta^\prime\ket{v_1,\uparrow}-\alpha^{\prime*}\ket{v_2,\downarrow}
	\end{align}
\end{subequations}
with:
\begin{subequations}
	\begin{align}
	\alpha^\prime&=\frac{2C_{v_1v_2}^*}{\left(4|C_{v_1v_2}|^2+W^{\prime 2}\right)^{1/2}} \\
	\beta^\prime&=\frac{W^\prime}{\left(4|C_{v_1v_2}|^2+W^{\prime 2}\right)^{1/2}}
	\end{align}
\end{subequations}
and:
\begin{equation}
W^\prime=\Delta-g\mu_B B+\sqrt{(\Delta-g\mu_B B)^2+4|C_{v_1v_2}|^2}\,.
\end{equation}
We can finally compute the Rabi frequencies for the resonant transitions between the ground-state $\ket{\psi_-}$ and the mixed spin and valley states $\ket{\psi_\pm^\prime}$:
\begin{subequations}
	\label{eqrabid}
	\begin{align}
	hf_-&=e\delta V_g\left|\bra{\psi_-^\prime}D\ket{\psi_-}\right|=e\delta V_g|\alpha^\prime\beta+\alpha^*\beta^\prime||D_{v_1v_2}| \label{eqrabim} \\
	hf_+&=e\delta V_g\left|\bra{\psi_+^\prime}D\ket{\psi_-}\right|=e\delta V_g|\alpha\alpha^\prime-\beta\beta^\prime||D_{v_1v_2}|\,.
	\end{align}
\end{subequations}
The Rabi frequencies for the resonant transitions between $\ket{\psi_+}$ and $\ket{\psi_\pm^\prime}$ are equivalent. We can also compute the Rabi frequency between the states $\ket{\psi^\prime_\pm}$:
\begin{equation}
hf_<=e\delta V_g\left|\bra{\psi_+^\prime}D\ket{\psi_-^\prime}\right|=e\delta V_g|\alpha^\prime\beta^\prime||D_{v_1v_1}-D_{v_2v_2}|\,,
\end{equation}
as well as the Rabi frequency between the states $\ket{\psi_\pm}$:
\begin{equation}
hf_>=e\delta V_g\left|\bra{\psi_+}D\ket{\psi_-}\right|=e\delta V_g|\alpha\beta||D_{v_1v_1}-D_{v_2v_2}|\,,
\end{equation}
where $D_{v_1v_1}=\bra{v_1,\sigma}D\ket{v_1,\sigma}$ and $D_{v_2v_2}=\bra{v_2,\sigma}D\ket{v_2,\sigma}$. The expansion of Eq. (\ref{eqrabim}) in powers of $B$ and $C_{v_1v_2}$ yields back Eq. (\ref{Seqrabipert}) at low magnetic fields. Yet Eqs. (\ref{eqrabid}) are valid up to much larger magnetic field (typically $g\mu_B B\lesssim E_3-E_1$, where $E_3$ is the energy of the next-lying state).

The typical energy diagram is plotted as a function of $B$ in Fig. 3b of the main text. From an experimental point of view, line A of Fig. 3b corresponds to $f_-$ ($B<B_{\rm V}$) or $f_+$ ($B>B_{\rm V}$), while line B corresponds to $f_<$. The line D on Fig. 3c would correspond to $f_>$.

\newpage
\section*{Supplementary note 3: Tight-binding modeling}
\label{sectionTB}

\subsection{Methodology and devices}
\label{sectionMethodology}

We have validated the model of supplementary note 2 against tight-binding (TB) calculations.\cite{SDiCarlo03,SDelerue05} TB is well suited to the description of such devices as it accounts for valley and spin-orbit coupling at the atomistic level (without the need for, e.g., extrinsic Rashba or Dresselhaus-like terms in the Hamiltonian).

\begin{figure}
	\centering
	\includegraphics[width=0.45\columnwidth]{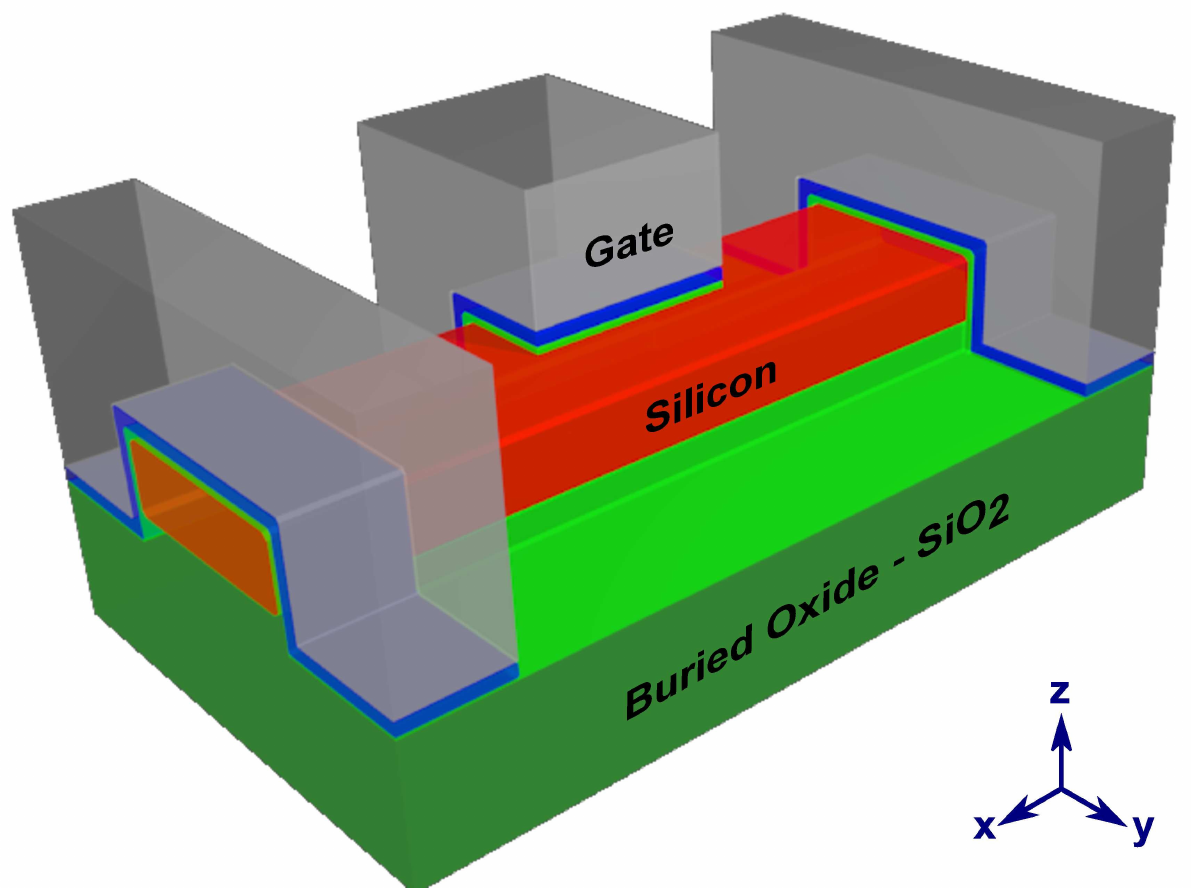}
	\hspace{1cm}
	\includegraphics[width=0.45\columnwidth]{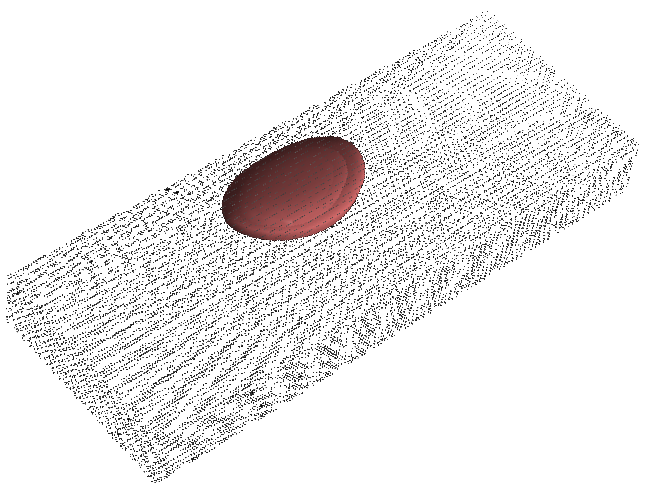}
	\caption{{\label{figdev} (left)} Electrostatic model of the device. The silicon wire appears in red, SiO$_2$ in green, HfO$_2$ in blue and the gate in gray. (right) Envelope of the squared TB wave function of the lowest-lying eigenstate ($V_g=0.1$ V).}
\end{figure}

We consider the prototypical device of Fig. \ref{figdev}a. The $[110]$ silicon nanowire (dielectric constant $\varepsilon_{\rm Si}=11.7$) is $W=30$ nm wide and $H=10$ nm thick. It is etched in a $(001)$ SOI film on top a 25 nm thick buried oxide (BOX).\footnote{The BOX used in the simulations is thinner than in the experiments (145 nm) for computational convenience. We have checked that this does not have any sizable influence on the results.} The QD is defined by the central, 30 nm long gate. This gate covers only part of the nanowire in order to confine a well-defined ``corner state''. The QD is surrounded by two lateral gates that control the barrier height between the dots (periodic boundary conditions being applied along the wire). The front gate stack is made of a layer of SiO$_2$ ($\varepsilon_{\rm SiO_2}=3.9$) and a layer of HfO$_2$ ($\varepsilon_{\rm HfO_2}=20$). The device is embedded in Si$_3$N$_4$ ($\varepsilon_{\rm Si_3N_4}=7.5$). We did not include the lateral gate in the simulations. All terminals are grounded except the central gate.

We compute the first four eigenstates $\ket{1}...\ket{4}$ of this device using a $sp^3d^5s^*$ TB model.\cite{SNiquet09} The dangling bonds at the surface of silicon are saturated with pseudo-hydrogen atoms. We include the effects of SOC and magnetic field. The SOC Hamiltonian is written as a sum of intra-atomic terms\cite{SChadi77} 
\begin{equation}
H_{\rm SOC}^{\rm TB}=2\lambda\sum_i\vec{L}_i\cdot\vec{S}\,,
\label{eqSOCTB}
\end{equation}
where $\vec{L}_i$ is the angular momentum on atom $i$, $\vec{S}$ is the spin and $\lambda$ is the SOC constant of silicon. The action of the magnetic field on the spin is described by the bare Zeeman Hamiltonian $H_{\rm z}=g_0\mu_B\vec{B}\cdot\vec{S}$, and the action of the magnetic field on the orbital motion of the electrons is accounted for by Peierl's substitution.\cite{SVogl95} We can then monitor the different Rabi frequencies
\begin{equation}
hf_{ij}=e\delta V_g\left|\bra{i}D\ket{j}\right|\,.
\end{equation}
The wave function of the ground-state $\ket{1}$ is plotted in Fig. \ref{figdev}b ($V_g=0.1$ V). It is, as expected for etched SOI structures, confined in a ``corner dot'' below the gate. Note that this TB description goes beyond the model of supplementary note 2 in including the action of the magnetic field on the orbital motion, and in dealing with all effects non-perturbatively. 

In the following, we define $x=[110]$ (nanowire axis), $y=[\bar{1}10]$ (perpendicular to the nanowire) and $z=[001]$ (perpendicular to the substrate).

\subsection{Dependence on the magnetic field amplitude}
\label{sectionMagneticFieldAmplitude}

\begin{figure}
	\centering
	\includegraphics[width=0.68\columnwidth]{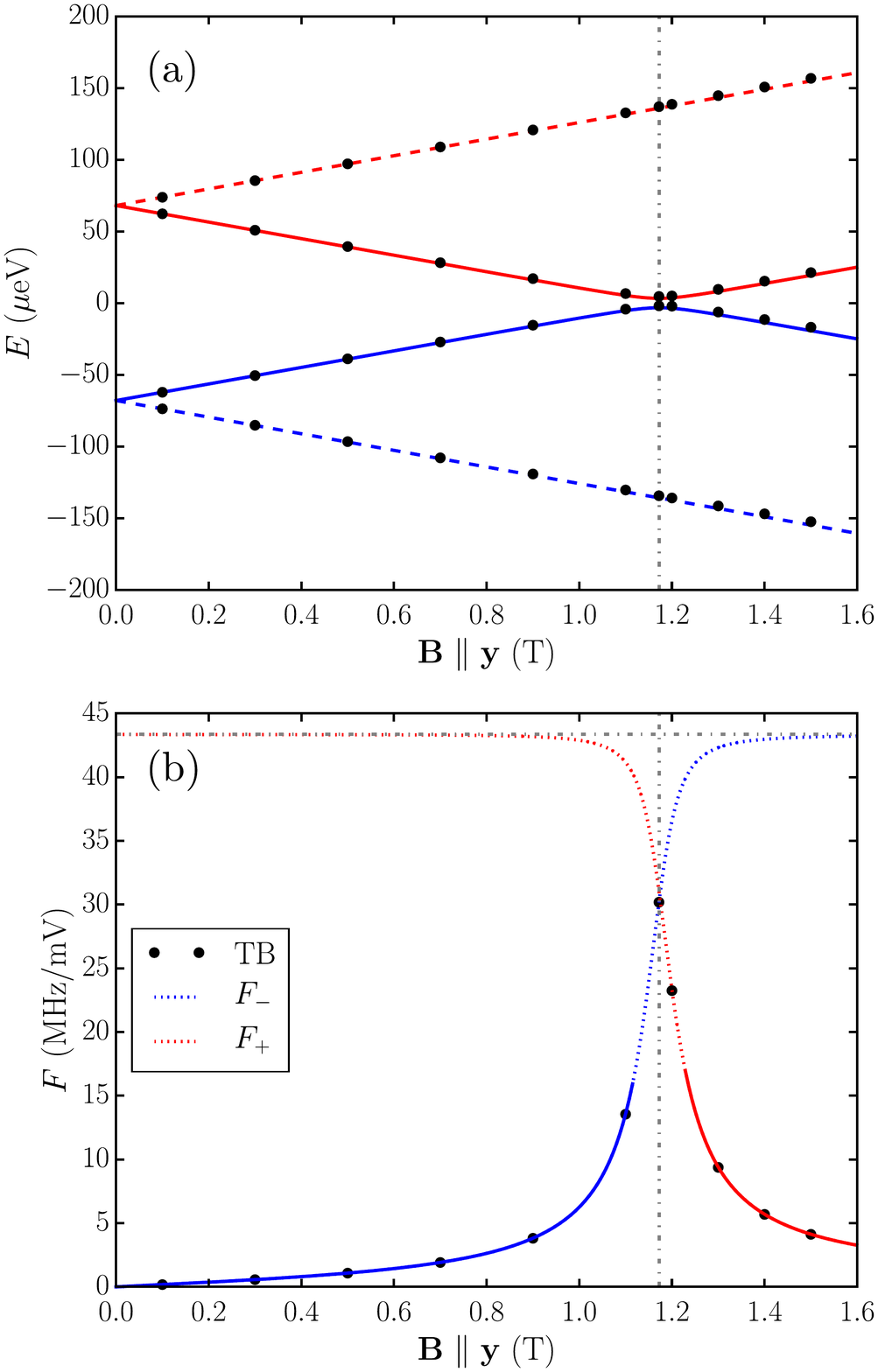}
	\caption{\label{fig_frabi_B} (a) Energy levels as a function of the magnetic field $\vec{B}\parallel\vec{y}$. (b) Reduced Rabi frequency $F=f/\delta V_g$ as a function of the magnetic field. The TB data are compared to the model of supplementary note 2. The vertical dash-dotted line is the anti-crossing field $B_{\rm V}=\Delta/(g\mu_B)$. The horizontal dash-dotted line is the maximal Rabi frequency $hF_{\rm max}=e|D_{v_1v_2}|$.}
\end{figure}

The energy of the first four eigenstates is plotted as a function of $\vec{B}\parallel\vec{y}$ in Fig. \ref{fig_frabi_B}a. States $\ket{1}$...$\ket{4}$ all belong to the $\Delta_z$ valleys, as confinement remains stronger along $z$ than along $y$ in a very wide range of gate voltages. The lowest eigenstate $\ket{1}$ can be identified with $\ket{\psi_-}\simeq\ket{v_1,\downarrow}$, the second one $\ket{2}$ and the third one $\ket{3}$ with $\ket{\psi^\prime_\pm}$ ($\ket{v_1,\uparrow}$/$\ket{v_2,\downarrow}$ anti-crossing at $B=B_{\rm V}=1.17$ T), and the fourth one $\ket{4}$ with $\ket{\psi_+}\simeq\ket{v_2,\uparrow}$.

The reduced TB Rabi frequency $F=f/\delta V_g$ between $\ket{v_1,\downarrow}$ and $\ket{v_1,\uparrow}$ is plotted as a function of $\vec{B}\parallel\vec{y}$ in Fig. \ref{fig_frabi_B}. Hence $F\equiv F_-$ before the anti-crossing between $\ket{v_1,\uparrow}$ and $\ket{v_2,\downarrow}$ at $B=B_{\rm V}$, and $F\equiv F_+$ after that anti-crossing. This transition corresponds to line A of the main text. The dependence of $F$ on $B$ is strongly non-linear, with a prominent peak near the anti-crossing. 

We can then switch off SOC, recompute the TB eigenstates at $\vec{B}=\vec{0}$, extract
\begin{subequations}
	\begin{align}
	\Delta&=E_2-E_1=135.86\ \mu{\rm eV} \\
	\left|C_{v_1v_2}\right|&=\left|\bra{v_2,\uparrow}H_{\rm SOC}\ket{v_1,\downarrow}\right|=3.25\ \mu{\rm eV} \\
	\left|D_{v_1v_2}\right|&=\left|\bra{v_1,\sigma}D\ket{v_2,\sigma}\right|=179.26\ \mu{\rm V/V} \\
	\left|D_{v_1v_1}-D_{v_2v_2}\right|&=\left|\bra{v_1,\sigma}D\ket{v_1,\sigma}-\bra{v_2,\sigma}D\ket{v_2,\sigma}\right|=813.26\ \mu{\rm V/V}
	\end{align}
\end{subequations}
and input these into Eqs. (\ref{eqrabid}). As shown in Fig. \ref{fig_frabi_B}, Eqs. (\ref{eqrabid}) perfectly reproduce the TB data. There are, in particular, no virtual transitions outside the lowest four $\ket{v_n,\sigma}$ that contribute significantly to the Rabi frequency.

\begin{figure}
	\centering
	\includegraphics[width=0.6\columnwidth]{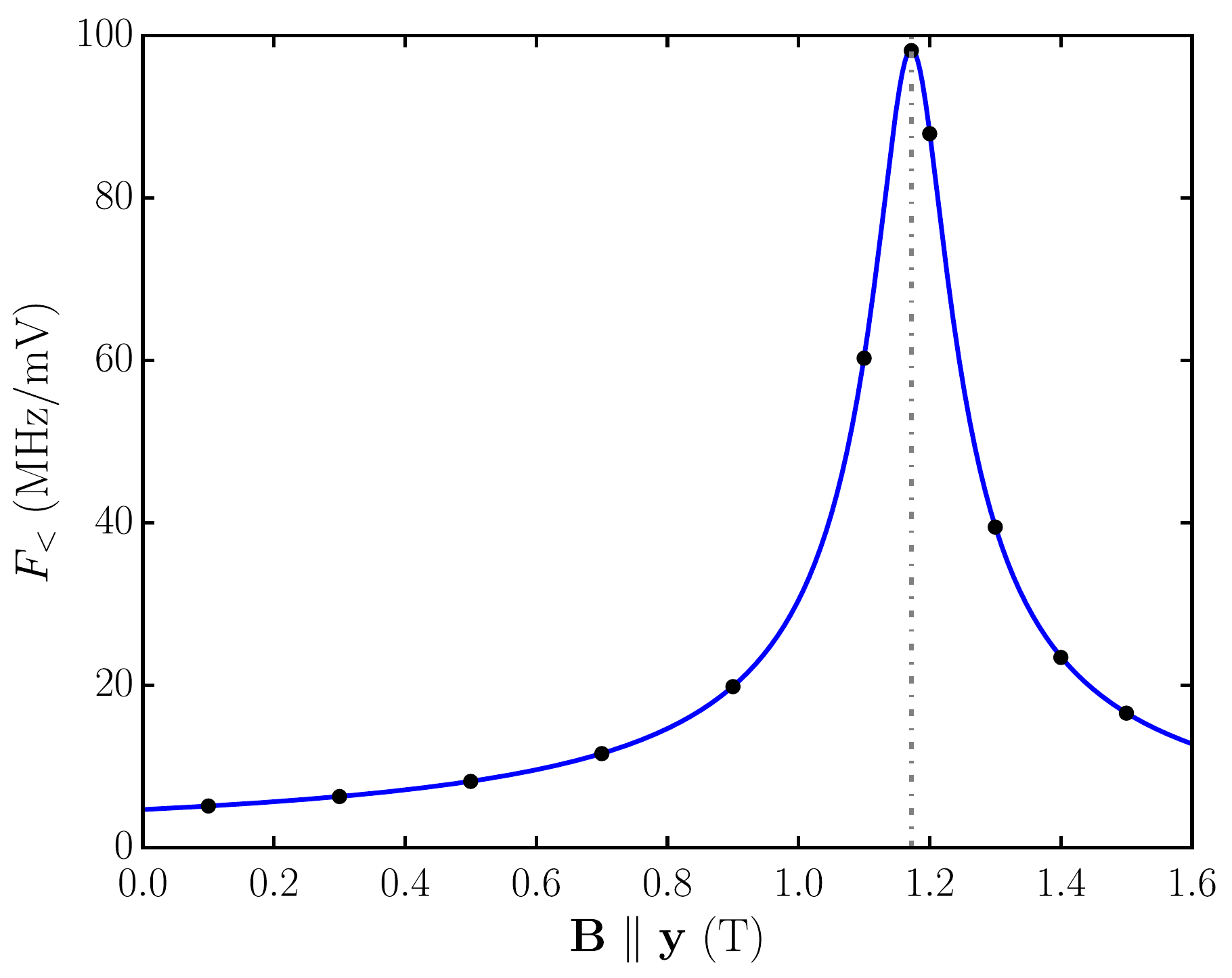}
	\caption{\label{fig_finf_B} Reduced Rabi frequency $F_<=f_</\delta V_g$ as a function of the magnetic field. The TB data (symbols) are compared to the model of supplementary note 2 (line). The vertical dash-dotted line is the anti-crossing field $B_{\rm V}=\Delta/(g\mu_B)$.}
\end{figure}

The reduced TB Rabi frequency $F_<=f_</\delta V_g$ is also plotted on Fig. \ref{fig_finf_B}. This transition corresponds to line B of the main text. It also shows a peak near the $B=B_{\rm V}$, and is pretty strong. 

In the present device, the TB valley splitting is much larger than the experiment ($\Delta\simeq 36\,\mu$eV). However, $\Delta$ decreases once surface roughness is introduced\cite{SCulcer10} and can range from a few tens to a few hundreds of $\mu$eV depending on the bias conditions. The TB data reported in Fig. 4 of the main text have been computed for a particular realization of surface roughness disorder that reproduces the experimental $\Delta\simeq 36\,\mu$eV. The surface roughness profiles used in these simulations have been generated from a Gaussian auto-correlation function with rms $\Delta_{\rm SR}=0.4$ nm and correlation length $\Lambda_{\rm SR}=1.5$ nm.\cite{SGoodnick85,SBourdet16} Note that disorder also reduces $|C_{v_1v_2}|$, $|D_{v_1v_2}|$, and $|D_{v_1v_1}-D_{v_2v_2}|$. A strategy for the control of the valley splitting in etched SOI devices will be reported elsewhere.

It is clear from Eqs. (\ref{eqrabid}) that $F_\pm<F_{\rm max}$, where $hF_{\rm max}=e|D_{v_1v_2}|$ is limited only by the dipole matrix element $D_{v_1v_2}$. In particular, near the anti-crossing field $B=B_{\rm V}$,
\begin{equation}
hF_+\simeq hF_-=F_{\rm ac}=\frac{e|D_{v_1v_2}|}{\sqrt{2}}
\end{equation}
only depends on $D_{v_1v_2}$. The SOC matrix element $C_{v_1v_2}$ actually controls the width of the peak around $B=B_{\rm V}$. The full width at half-maximum $\Delta B_{\rm FWHM}$ of this peak ($F=F_{\rm ac}/2$) indeed reads:
\begin{equation}
g\mu_B\Delta B_{\rm FWHM}\simeq\frac{12|C_{v_1v_2}|}{\sqrt{7}}\,.
\end{equation}

\newpage
\section*{Supplementary note 4: Role of symmetries}
\label{sectionRoleofSymmetries}

\begin{figure}[!h]
	\centering
	\includegraphics[width=0.45\columnwidth]{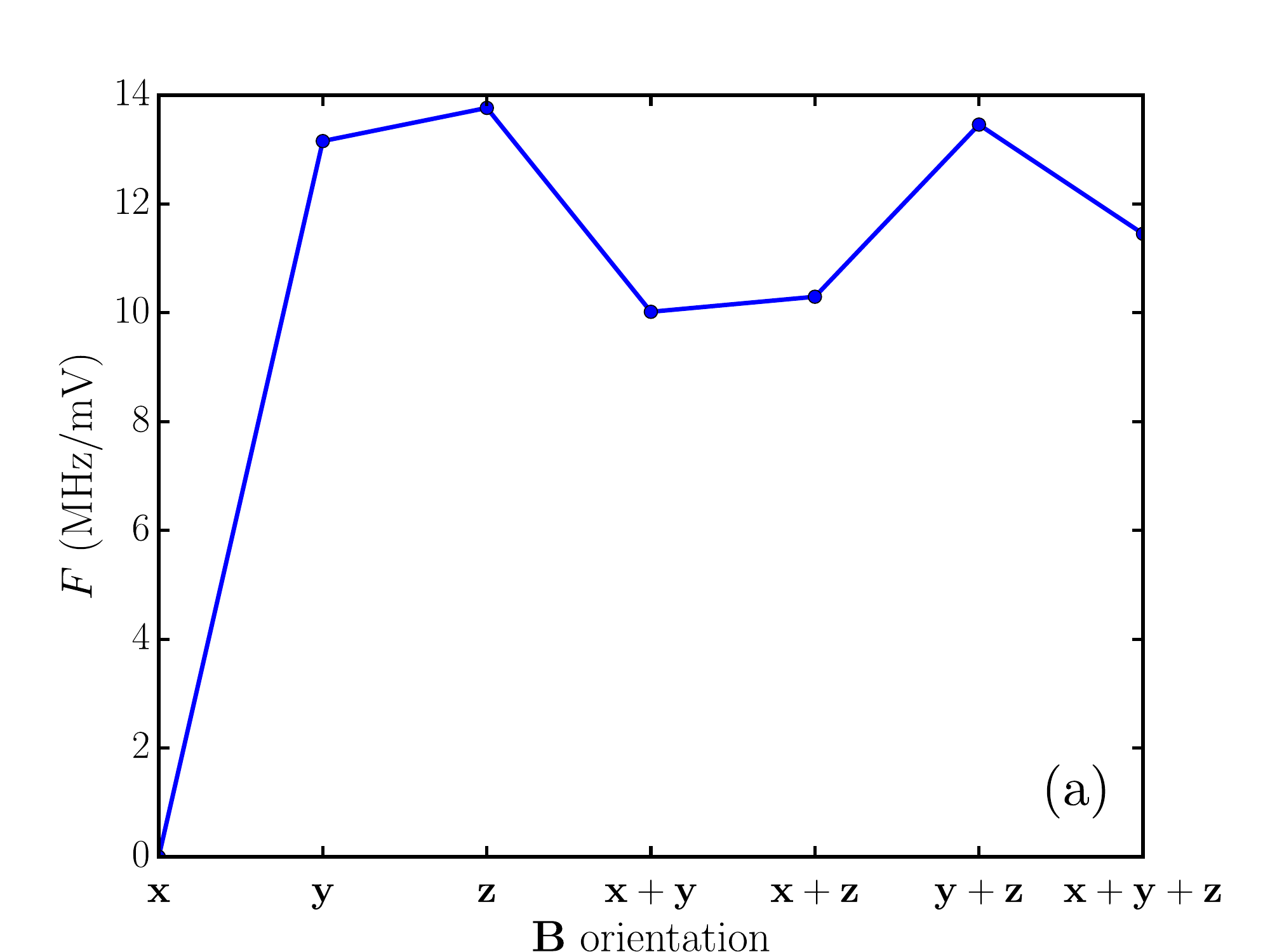}
	\includegraphics[width=0.45\columnwidth]{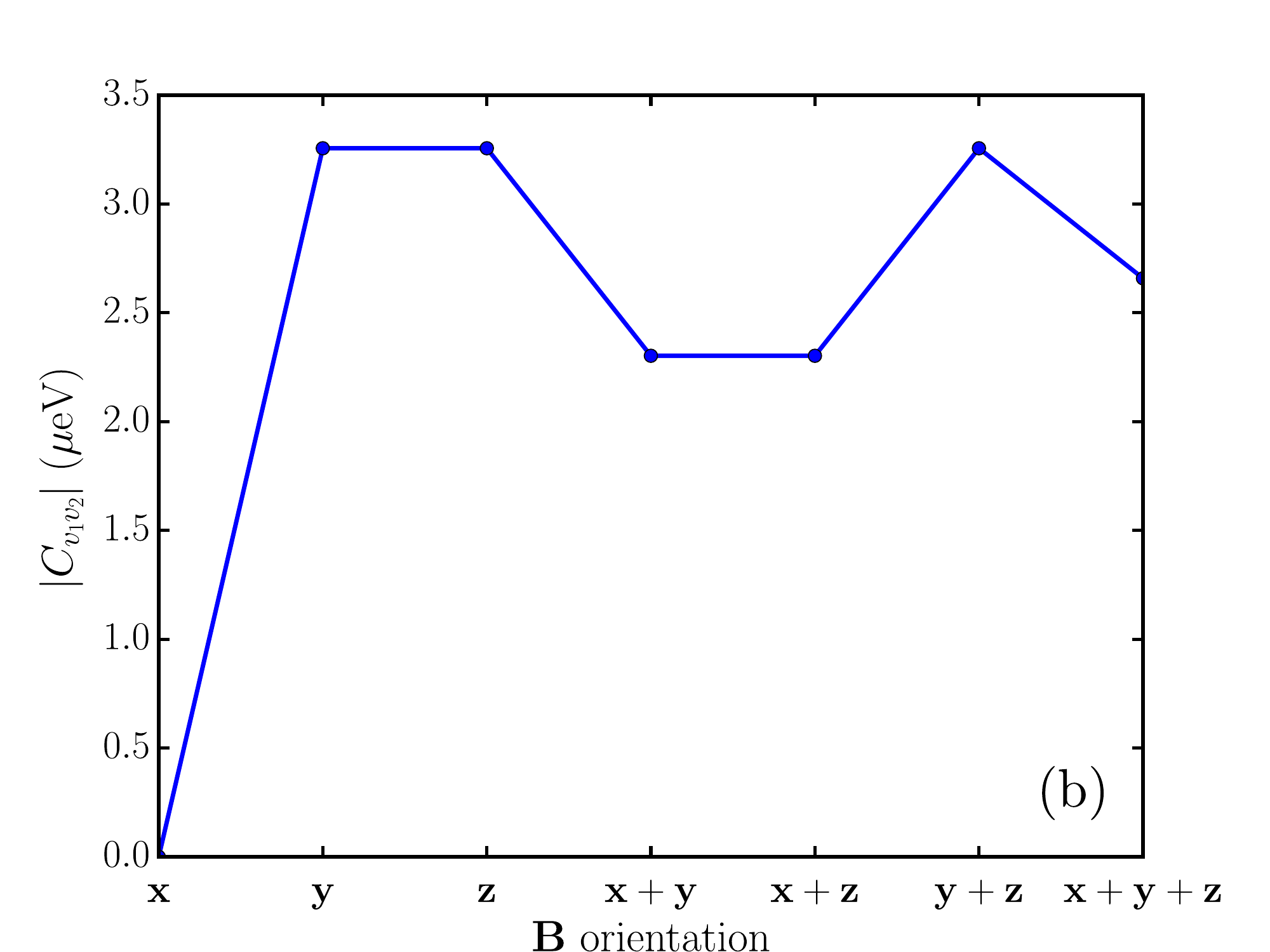}
	\caption{(a) Reduced Rabi frequency $F=f/\delta V_g$ and (b) SOC matrix element $|C_{v_1v_2}|$ as a function of the magnetic field orientation ($|\vec{B}|=1.1$ T).\label{fig_frabi_Bor}}
\end{figure}

To highlight the role of symmetries, we plot the reduced TB Rabi frequency $F$ (corresponding to line A of the main text) and SOC matrix element $|C_{v_1v_2}|$ as a function of the orientation of the magnetic field in Fig. \ref{fig_frabi_Bor} (same bias conditions as in Fig. \ref{fig_frabi_B}, $|\vec{B}|=1.1$ T). As expected, $F$ shows the same trends as $|C_{v_1v_2}|$ -- although there are small discrepancies. These discrepancies result from the effects of the magnetic field on the orbital motion of the electron, dismissed in the supplementary note 2. 

\begin{figure}
	\begin{center}
		\includegraphics[width=0.45\columnwidth]{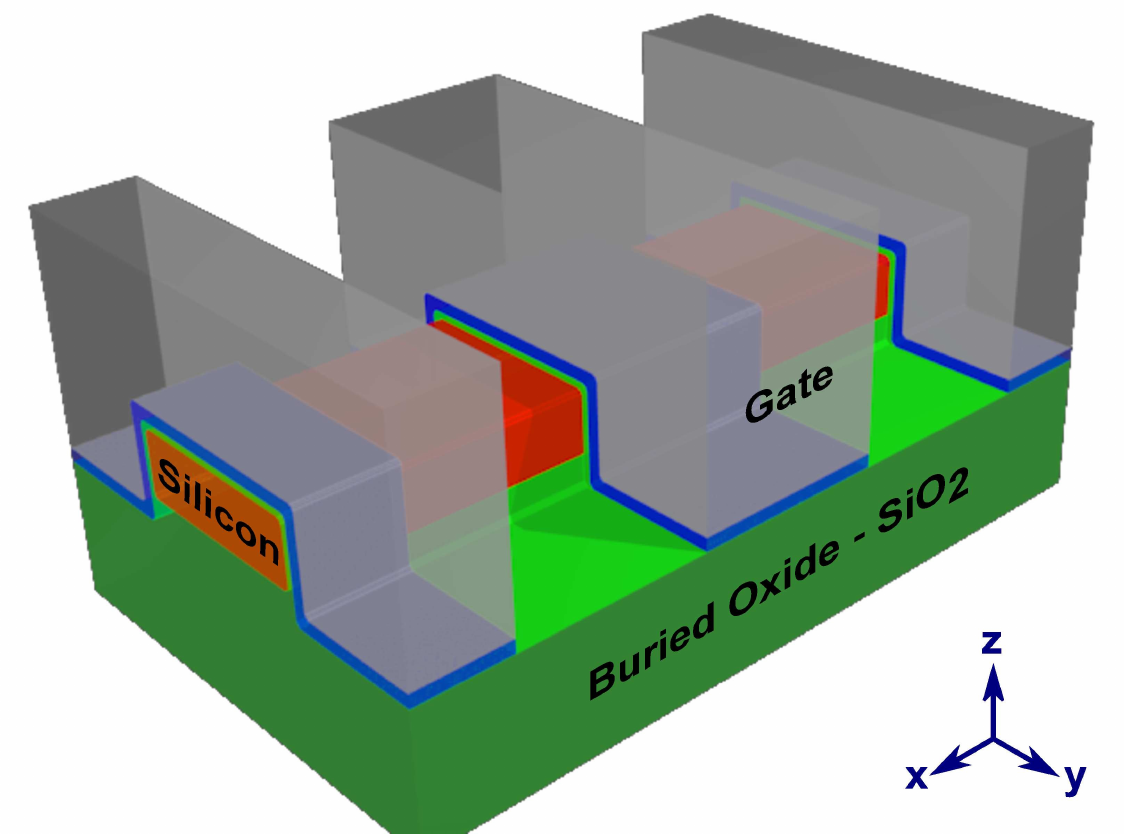}
		\hspace{1cm}
		\includegraphics[width=0.45\columnwidth]{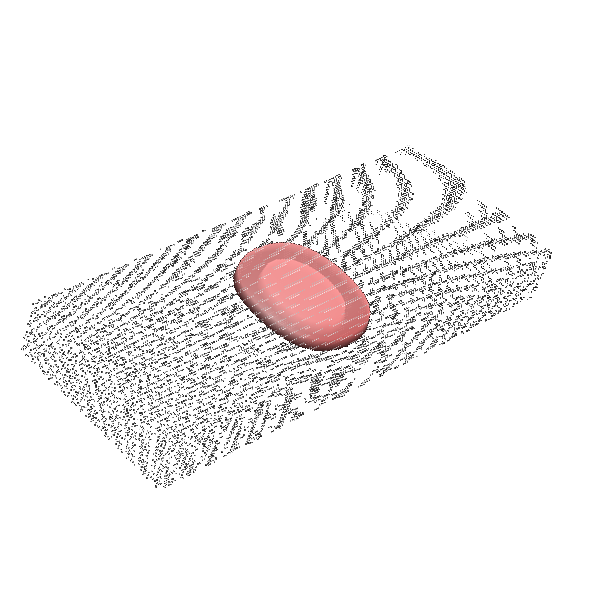}
		\caption{\label{fig_wfn_topgate} (left) Electrostatic model of the trigate device. At variance with Fig. \ref{figdev}, the central gate covers all three sides of the nanowire. (right) Envelope of the squared TB wave function of the lowest-lying eigenstate ($V_g=0.1$ V).}
	\end{center}
\end{figure}

Strikingly, the SOC matrix element and Rabi frequency are almost zero when the magnetic field is aligned with the nanowire axis ($x$). The dependence of $C_{v_1v_2}$ on the magnetic field (or spin) orientation suggests that only the $\propto\sigma_x$ term of the SOC Hamiltonian is relevant for this matrix element. This can be supported by a symmetry analysis. Now quantifying the spin along $z$, the TB SOC Hamiltonian [Eq. (\ref{eqSOCTB})] reads:
\begin{equation}
H_{\rm SOC}^{\rm TB}=\lambda\left({\cal L}_x\sigma_x+{\cal L}_y\sigma_y+{\cal L}_z\sigma_z\right)\,,
\end{equation}
where $\sigma_\alpha$ are the Pauli matrices, ${\cal L}_\alpha=\sum_iL_{i,\alpha}$ and $L_{i,\alpha}$ is the component 
$\alpha=x,y,z$ of the angular momentum on atom $i$. The device of Fig. \ref{fig_wfn_topgate} is only invariant by reflection through the $(yz)$ plane (space group $C_s$). The table of characters of this space group is reproduced in Table \ref{tabgroups}. Let us remind that for any observable $O$, unitary transform (symmetry operation) $R$, and wave functions $\ket{\psi_1}$ and $\ket{\psi_2}$,
\begin{equation}
O_{12}=\bra{\psi_1}O\ket{\psi_2}=\bra{R\psi_1}ROR^\dagger\ket{R\psi_2}
\label{eqsymmetry}
\end{equation}

\begin{table}
	\centering
	\begin{tabular}{l|r|r}
		$C_s$ & $E$ & $\sigma(yz)$ \\
		\hline
		$A_1$ & $1$ & $+1$         \\
		$A_2$ & $1$ & $-1$              
	\end{tabular}
	\hspace{1 cm}
	\begin{tabular}{l|r|r|r|r}
		$C_{2v}$ & $E$ & $\sigma(yz)$ & $\sigma(xz)$ & $C_2(z)$ \\
		\hline
		$A_1$    & $1$ & $+1$         & $+1$         & $+1$     \\
		$A_2$    & $1$ & $-1$         & $-1$         & $+1$     \\
		$B_1$    & $1$ & $+1$         & $-1$         & $-1$     \\
		$B_2$    & $1$ & $-1$         & $+1$         & $-1$     \\
	\end{tabular}
	\caption{\label{tabgroups}Table of characters of the $C_s$ and $C_{2v}$ groups.}
\end{table}

\begin{table}
	\centering
	\begin{tabular}{ccc}
		$L_x=
		\begin{spmatrix}{}
		0 & 0 &  0 \\   
		0 & 0 & -i \\   
		0 & i &  0 
		\end{spmatrix}$
		&
		$L_y=
		\begin{spmatrix}{}
		0 & 0 & i \\   
		0 & 0 & 0 \\   
		-i & 0 & 0 
		\end{spmatrix}$
		&
		$L_z=
		\begin{spmatrix}{}
		0 & -i & 0 \\   
		i &  0 & 0 \\   
		0 &  0 & 0 
		\end{spmatrix}$
	\end{tabular}
	\caption{\label{tabL}Matrices of $L_x$, $L_y$ and $L_z$ in the atomic $\{p_x, p_y, p_z\}$ basis set.}
\end{table}

Both $\ket{v_1}$ and $\ket{v_2}$ states belong to the irreducible representation $A_1$ of $C_s$. Therefore, $\sigma(yz)\ket{v_1}=\ket{v_1}$ and $\sigma(yz)\ket{v_2}=\ket{v_2}$. Also, $D(\vec{r})=\partial V_t(\vec{r})/\partial V_g$ is compatible with the $C_s$ symmetry, so that $\sigma(yz)D\sigma(yz)^\dagger=D$. Therefore, Eq. (\ref{eqsymmetry}) does not set any condition on $D_{v_1v_2}$. However, $\sigma(yz)$ transforms a $p_{x}$ orbital into a $-p_{x}$ orbital, hence transforms ${\cal L}_{y}$ into $-{\cal L}_{y}$, ${\cal L}_z$ into $-{\cal L}_z$, but leaves ${\cal L}_x$ invariant (see Table \ref{tabL}). Eq. (\ref{eqsymmetry}) then imposes that ${\cal L}_y$ and ${\cal L}_z$ can not couple the $\ket{v_1}$ and $\ket{v_2}$ states. This is why only the $\propto{\cal L}_x\sigma_x$ term of the atomistic SOC Hamiltonian makes a non-zero contribution. As a result, $|C_{v_1v_2}|$ and the Rabi frequency are proportional to $|\sin\theta|$, where $\theta$ is the angle between $\vec{x}$ and an in-plane magnetic field. The effective SOC Hamiltonian between $\ket{v_1,\downarrow}$ and $\ket{v_2,\uparrow}$ is likely dominated by Rashba-type contributions of the form $H_{\rm SOC}^{\rm eff}\propto p_{y,z}\sigma_x$ (where $p_y$ and $p_z$ are the momenta along $y$ and $z$). These Rashba-type contributions may arise from the electric field of the gate along $y$ and $z$, and/or from the top and lateral Si/SiO$_2$ interfaces bounding the corner dot.

How far is the EDSR connected with the formation of corner dots in etched SOI structures ? To answer this question, we have computed the Rabi frequency in a standard ``Trigate'' device where the gate surrounds the channel on three sides (see Fig. \ref{fig_wfn_topgate}a). The dimensions are the same as in Fig. \ref{figdev}, but the central gate now covers the totality of  the nanowire. As shown in Fig. \ref{fig_wfn_topgate}b, the low-lying conduction band states are still confined at the top interface, but the wave function is symmetric with respect to the $(xz)$ plane (no ``corner'' effect at zero back gate voltage). It turns out that there is no EDSR whatever the orientation of the magnetic field. Actually, the SOC matrix element $C_{v_1v_2}$ is zero for all spin orientations.

Therefore the confinement in a corner dot seems to be a pre-requisite for a sizable EDSR. As a matter of fact, the devices of Figs. \ref{fig_wfn_topgate} and \ref{figdev} have different symmetries. Indeed, the device of Fig. \ref{fig_wfn_topgate} is invariant by reflections through the $(yz)$ and $(xz)$ planes, and by the twofold rotation around the $z$ axis (space group $C_{2v}$). The states $\ket{v_1}$ and $\ket{v_2}$ of this device again belong to the irreducible representation $A_1$ of $C_{2v}$ (Table \ref{tabgroups}). For the same reasons as before,  Eq. (\ref{eqsymmetry}) does not set any condition on $D_{v_1v_2}$. The existence of a $\sigma(yz)$ mirror still imposes that only ${\cal L}_x$ can couple $\ket{v_1}$ and $\ket{v_2}$ states. Yet the introduction of a $\sigma(xz)$ mirror, which leaves only ${\cal L}_y$ invariant, prevents such a coupling. Therefore, none of the operators ${\cal L}_x$, ${\cal L}_y$ and ${\cal L}_z$ can couple the $\ket{v_1}$ and $\ket{v_2}$ states. $C_{v_1v_2}$ is thus zero, and SOC-mediated EDSR is not possible.

To conclude, the symmetry must be sufficiently low in order to achieve EDSR in the conduction band of silicon. This condition is realized in corner dots where there is only one $\sigma(yz)$ mirror left (yet preventing EDSR for a magnetic field along the wire axis). The design of geometries without any symmetry left would in principle maximize the opportunities for EDSR.

\end{document}